\documentclass[12pt]{article}
\usepackage{epsfig, times, fullpage, setspace, array, psfig}
\usepackage{subfigure, color, amsmath, url, amsfonts, amssymb}
\usepackage{multicol,float,multirow,colortbl}
\usepackage{rotating}
\newtheorem{thm}{Theorem}[section]
\newtheorem{lem}[thm]{Lemma}
\newtheorem{cor}[thm]{Corollary}

\newtheorem{den}[thm]{Definition}

\def\FullBox{\hbox{\vrule width 8pt height 8pt depth 0pt}}

\def\qed{\ifmmode\qquad\FullBox\else{\unskip\nobreak\hfil
\penalty50\hskip1em\null\nobreak\hfil\FullBox
\parfillskip=0pt\finalhyphendemerits=0\endgraf}\fi}

\title{Approximating the Minimum Breakpoint Linearization Problem for Genetic Maps without Gene Strandedness}
\author{Xin Chen\\
Division of Mathematical Sciences \\
School of Physical and Mathematical Sciences \\
Nanyang Technological University, Singapore}

\date{}
\begin{document}
\maketitle

\begin{abstract}
The study of genetic map linearization leads to a combinatorial
hard problem, called the {\em minimum breakpoint linearization}
(MBL) problem. It is aimed at finding a linearization of a partial
order which attains the minimum breakpoint distance to a reference
total order. The approximation algorithms previously developed for
the MBL problem are only applicable to genetic maps in which genes
or markers are represented as signed integers. However, current
genetic mapping techniques generally do not specify gene
strandedness so that genes can only be represented as unsigned
integers. In this paper, we study the MBL problem in the latter
more realistic case. An approximation algorithm is thus developed,
which achieves a ratio of $(m^2+2m-1)$ and runs in $O(n^7)$ time,
where $m$ is the number of genetic maps used to construct the
input partial order and $n$ the total number of distinct genes in
these maps.
\end{abstract}


\section{Introduction}

Genetic map linearization is a crucial preliminary step to most
comparative genomics studies, because they generally require a
total order of genes or markers on a chromosome rather than a
partial order that current genetic mapping techniques might only
suffice to provide \cite{BB07,FJ07,ZL05,ZS06}. One of the
computational approaches proposed for genetic map linearization is
to find a topological sort of the directed acyclic graph (DAG)
that represents the input genetic maps while minimizing its
breakpoint distance to a reference total order. It hence leads to
a combinatorial optimization problem, called the {\em minimum
breakpoint linearization} (MBL) problem \cite{BB07}, which has
attracted great research attention in the past few years
\cite{BB07,BF10,CC09,FJ07}.

The MBL problem is already shown to be {\bf NP}-hard \cite{BB07},
and even {\bf APX}-hard \cite{BF10}. The first algorithm proposed
to solve the MBL problem is an exact dynamic programming algorithm
running in exponential time in the worst case \cite{BB07}. In the
same paper, a time-efficient heuristic algorithm is also
presented, which, however, has no performance guarantee. The first
attempt was made in \cite{CC09} to develop a polynomial-time
approximation algorithm. Unfortunately, the proposed algorithm was
latter found invalid \cite{BF10} because it relies on a flawed
statement in \cite{FJ07} on {\em adjacency-order graphs}. To fix
this flaw, the authors of \cite{BF10} revised the construction of
adjacency-order graphs and proposed three approximation
algorithms, two of which are based on the existing approximation
algorithms for a general variant of the {\em feedback vertex set}
problem, and the third was instead developed in the same spirit as
was done in \cite{CC09}, achieving a ratio of $(m^2+4m-4)$ (only
for $m\ge 2$).

As we shall show in Section~\ref{sec:AOG}, the above approximation
algorithms are only applicable to the input genetic maps in which
genes or markers are represented as signed integers, where the
signs represent the strands of genes/markers. However, we note
that the original definition of the MBL problem assumes unsigned
integers for genes \cite{BB07}. In fact, this is a more realistic
case. Current genetic mapping techniques such as recombination
analysis and physical imaging generally do not specify gene
strandedness so that genes can only be represented as unsigned
integers \cite{ZL05}. Based on this observation, whether the MBL
problem can be approximated still remains a question not yet to be
resolved.

In this paper, we study the MBL problem in the more realistic case
where no gene strandedness information is available for the input
genetic maps. We revised the definition of conflict-cycle in
\cite{BF10}, from which an approximation algorithm is hence
developed also in the same spirit as done in \cite{BF10, CC09}. It
achieves a ratio of $(m^2+2m-1)$ (which holds for all $m\ge 1$)
and runs in $O(n^7)$ time, where $m$ is the number of genetic maps
used to construct the input partial order and $n$ the total number
of distinct genes occurring in these maps.

The rest of the paper is organized as follows. We first introduce
some preliminaries and notations in Section~\ref{sec:preliminary}.
In Section~\ref{sec:facts} we discuss a number of basic facts
about the MBL problem, which leads to the formulation of the {\em
minimum breakpoint vertex set} (MBVS) problem in
Section~\ref{sec:approximation}. We present an approximation
algorithm for the MBL problem via the approximation of the MBVS
problem in Section~\ref{sec:approximation}, and then conduct
performance analyses on both its approximation ratio and running
time in Section~\ref{sec:performance}. Finally, some concluding
remarks are made in Section~\ref{sec:conclusion}. For the sake of
consistency, we borrowed many notations from \cite{BF10} and
\cite{CC09} throughout the paper.

\section{Preliminaries and notations} \label{sec:preliminary}

\subsection{Genetic maps and their combined directed acyclic graph}
\label{sec:gene-map}

A genetic map is a totally-ordered sequence of {\em blocks}, each
of which comprises one or more genes. It defines a partial order
on genes, where genes within a block are ordered before all those
in its succeeding blocks, but unordered among themselves.

\begin{figure*}[t]
\begin{center}
  \includegraphics[width=0.9\textwidth]{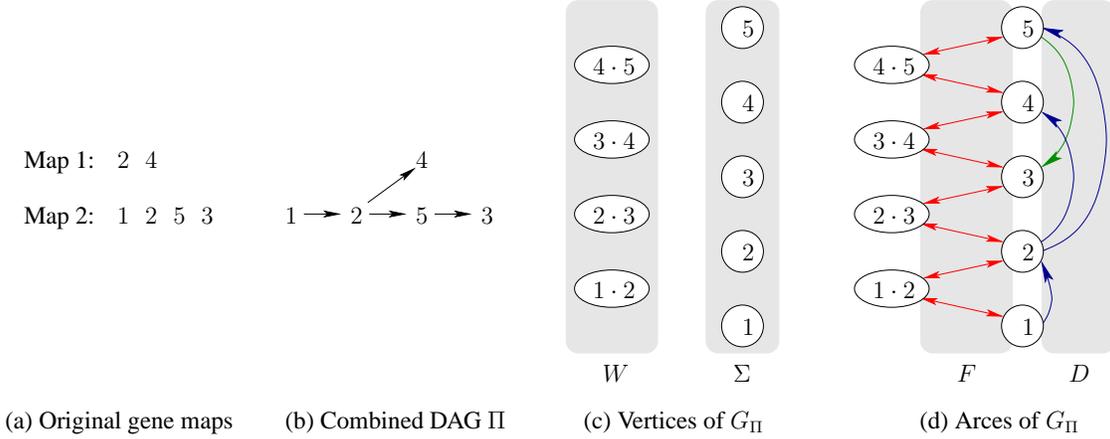}
\caption{The construction of an adjacency-order graph as proposed
in \cite{BF10}. The symmetric arcs in $F$ are represented as
double arrows.} \label{fig:An-adjacency-graph}
\end{center}
\end{figure*}

Today it is increasingly common to find multiple genetic maps
available for a same genome. Combining these maps often provides a
partial order with a higher coverage of gene ordering than an
individual genetic map. To represent this partial order, we may
construct a directed acyclic graph $\Pi=(\Sigma, D)$, where the
vertex set $\Sigma = \{ 1, \dots, n\}$ is made of all the
contributing genes and the arc set $D$ made of all the ordered
pairs of genes appearing in consecutive blocks of the same genetic
map \cite{YS03, ZL05}. Two properties can be deduced \cite{BF10}
from these genetic maps and their combined directed acyclic graph:
(i) if there is an arc between two genes $i$ and $j$ in $\Pi$,
then $i$ and $j$ appear in consecutive blocks of some genetic map,
and (ii) if $i$ and $j$ appear in different blocks of the same
genetic map, then there exists in $\Pi$ a nonempty directed path
either from $i$ to $j$ or from $j$ to $i$. See
Figure~\ref{fig:An-adjacency-graph} for a simple example of $\Pi$
constructed from two genetic maps.

We say gene $i$ is ordered {\em before} (resp. {\em after}) gene
$j$ by $\Pi$ if there exists in $\Pi$ a nonempty directed path
from $i$ to $j$ (resp. $j$ to $i$). We use $i \prec_\Pi j$ to
denote the ordering relation that gene $i$ is ordered before gene
$j$ by $\Pi$. Unlike in \cite{CY10}, we assume in this paper that
combining multiple genetic maps would never create order
conflicts, i.e., we could not have both $i \prec_\Pi j$ and $j
\prec_\Pi i$ simultaneously.

\subsection{The minimum breakpoint linearization problem}

Let $\Pi = (\Sigma, D)$ be a directed acyclic graph representing a
partial order generated with $m$ genetic maps of a same genome. A
{\em linearization} of $\Pi$ is a total order of genes $\pi =
\pi(1)\cdot \pi(2) \cdots \pi(n)$, i.e., a permutation on
$\{1,2,\dots,n\}$, such that, for all genes $i,j$, if $i\prec_\Pi
j$ then $i\prec_\pi j$. In this case, $\pi$ is said to be {\em
compatible} with $\Pi$. Let $\Gamma$ denote another genome with
the same set of genes in a total order. Without loss of
generality, we assume that $\Gamma$ is the identity permutation
$1\ 2\ \cdots \ n$. A pair of genes that are adjacent in $\pi$ but
not in $\Gamma$ is called a {\em breakpoint} of $\pi$ with respect
to $\Gamma$, and the total number of breakpoints is thus defined
as the {\em breakpoint distance} between $\pi$ and $\Gamma$
\cite{BP93}.

Given a partial order $\Pi$ and a total order $\Gamma$ as
described above, the minimum breakpoint linearization (MBL)
problem is defined as to find a linearization $\pi$ of $\Pi$ such
that the breakpoint distance between $\pi$ and $\Gamma$ is
minimized \cite{BB07}. This minimum breakpoint distance is further
referred to as the breakpoint distance between $\Pi$ and $\Gamma$,
and denoted by $d_b(\Pi,\Gamma)$.

\subsection{Adjacency-Order Graph} \label{sec:AOG}

In this study we adopt the definition of adjacency-order graph
introduced in \cite{BF10}. To construct an adjacency-order graph
for a partial order $\Pi = (\Sigma, D)$, we first create a set $W$
of vertices representing the {\em adjacencies} of the identity
permutation $\Gamma$ by $W=\{i \cdot (i+1) | 1\leq i < n\}$, and
let $V=\Sigma \cup W$ (see Figure~\ref{fig:An-adjacency-graph}c).
We will not distinguish the vertices of $\Sigma$ and their
corresponding integers, which is always be clear from the context.
Then, we construct a set of arcs $F$ as
$$ F \ = \ \{ i\cdot(i+1) \to i,\ i\cdot(i+1) \to i+1,\
   i \to i \cdot(i+1),\ i+1 \to i \cdot(i+1)\ |\ 1\leq i < n\},$$
where the arrow $\to$ is used to denote an arc. Note that every
arc in $F$ has one end in $W$ and the other end in $\Sigma$. Let
$E=D\cup F$ (see Figure~\ref{fig:An-adjacency-graph}d). Finally,
we define the {\em adjacency-order} graph $G_\Pi$ of $\Pi$ by
$G_\Pi = (V, E)$.

Note that in $G_\Pi$, the arcs of $D$ may go either top-down or
bottom-up. Let $X[G_\Pi]$ (or only $X$, if there is no ambiguity)
be the set of arcs in $D$ that go top-down, and $Y[G_\Pi]$ (or
only $Y$) the set of arcs in $D$ that go bottom-up. Formally, we
may write $X[G_\Pi]=\{ i\to j \in D\ | \ i>j \}$ and $Y[G_\Pi]=\{
i\to j \in D\ | \ i<j \}$. It is easy to see that $D=X\cup Y$ and
$X\cap Y= \emptyset$.

In \cite{BF10}, a  conflict-cycle refers to a cycle that uses an
arc from $X$. By this definition, a conflict-cycle may not
necessarily use any arc from $Y$ and all its adjacencies might
still co-exist in some linearization of $\Pi$, as we can see from
the adjacency-order graph $G_\Pi$ shown in
Figure~\ref{fig:An-adjacency-graph}d. This adjacency-order graph
contains a conflict-cycle $3\to 3\cdot 4 \to 4 \to 4\cdot 5\to 5
\to 3$ (as defined in \cite{BF10}), for which both adjacencies
$3\cdot 4$ and $4\cdot 5$ may occur in the linearization $1\ 2\ 5\
4\ 3$ of $\Pi$. Based on these observations, in this study we use
a different definition of conflict-cycles as follows.
\begin{den}
A cycle in $G_\Pi$ is called a conflict-cycle if it contains at
least one arc from $X$ and at least one arc from $Y$.
\end{den}
This new definition has wide implications for the future
approximation of the MBL problem, as we shall see latter. A quick
look indicates that the example cycle mentioned above is no longer
a conflict-cycle. In Theorem~\ref{thm:conflict-cycle}, we shall
prove that the adjacencies involved in a conflict-cycle could not
co-exist in any linearization of $\Pi$. Consequently, we need to
remove at least one adjacency from each of those cycles in order
to obtain a linearization of $\Pi$.

Most of the following notations are already introduced in
\cite{BF10}. An arc between $u$ and $v$ is written $u\to v$, or
$u\to_A v$ if it belongs to some set $A$. A {\em path} $P$ is a
(possibly empty) sequence of arcs written $u \xrightarrow{P}{}^*
v$, or $u \xrightarrow{P}{}_A^* v$ if $P$ uses arcs only from $A$.
A nonempty path $Q$ is written as $u \xrightarrow{Q}{}^+ v$ with a
$+$ sign. A {\em cycle} is a nonempty path $u
\xrightarrow{\mathcal{C}}{}^+ v$ with $v=u$. Given a path $P= v_0
\to v_1 \to \dots \to v_l$ in $G_\Pi$, the following notations are
used: $l(P) = l$ is the length of $P$, $V(P) = \{ v_h\ | \ 0\leq
h\leq l\}$, $W(P) = V(P) \cap W$, $\Sigma (P) = V(P) \cap \Sigma$,
$E(P) = \{ v_h\to v_{h+1}\ | \ 0\leq h < l\}$, $F(P) = E(P) \cap
F$, $D(P) = E(P) \cap D$, $X(P) = E(P) \cap X$, and $Y(P) = E(P)
\cap Y$. A cycle $\mathcal{C}$ is said to be {\em simple} if all
vertices $v_h$ are distinct except $v_0 = v_h$, which implies that
$l(\mathcal{C}) = |V(\mathcal{C})| = |E(\mathcal{C})|$. If a cycle
$\mathcal{C}$ is not simple, then it contains a {\em subcycle}
$\mathcal{C}^{'}$ such that $V(\mathcal{C}^{'})\subseteq
V(\mathcal{C})$ and $E(\mathcal{C}^{'})\subseteq E(\mathcal{C})$.
In this paper, we further require $\mathcal{C}^{'}\ne \mathcal{C}$
when $\mathcal{C}^{'}$ is the subcycle of $\mathcal{C}$.

\section{Some basic facts} \label{sec:facts}

Given a cycle $\mathcal{C}$ in $G_\Pi$, we may partition
$W(\mathcal{C})$ into a collection of disjoint subsets
$W_h(\mathcal{C})$ such that each of them can be written as $\{
i\cdot (i+1) \ | \ a_h\leq i < b_h\}$, for some integers $a_h$ and
$b_h$. We denote such a collection of disjoint subsets with
minimum cardinality by $\mathbb{W}(\mathcal{C}) = \{
W_1(\mathcal{C}), W_2(\mathcal{C}), \cdots, W_l(\mathcal{C}) \}$.
Note that, for every cycle $\mathcal{C}$ in $G_\Pi$, we have
$l=|\mathbb{W} (\mathcal{C}) | \ge 1$ because $\Pi = (\Sigma, D)$
is a directed acyclic graph.

\begin{lem} \label{lem:106}
Let $\mathcal{C}$ be a (not necessarily simple) cycle with
$W_1(\mathcal{C}) = \{ i\cdot (i+1) \ | \ a_1\leq i < b_1\}$ and
$W_2(\mathcal{C}) = \{ i\cdot (i+1) \ | \ a_2\leq i < b_2\}$ being
two distinct elements of $\mathbb{W}(\mathcal{C})$. Then, we have
$[a_1, b_1] \cap [a_2, b_2] = \emptyset$.
\end{lem}

\noindent {\em Proof.} \indent By contradiction, suppose that
$[a_1, b_1] \cap [a_2, b_2] \ne \emptyset$, which implies that
$a_1\le b_2$ and $a_2\le b_1$. Let $a=\min (a_1, a_2)$ and $b=\max
(b_1, b_2)$, and let $W_1^{'}(\mathcal{C}) = \{ i\cdot (i+1) \ | \
a\leq i < b\}$. For $\forall i\in [a_1,b_1] \cup [a_2,b_2]$, we
have $i\in [a,b]$, which implies that $W_1(\mathcal{C}) \cup
W_2(\mathcal{C}) \subseteq W_1^{'}(\mathcal{C})$. Next we show
that, for $\forall i\in [a,b)$, we have either $i\in [a_1, b_1)$
or $i\in [a_2, b_2)$. If $i\notin [a_1, b_1)$, then $i\ge b_1$
since $i\ge a_1$ and, further, $i\ge a_2$ since $a_2\le b_1$. On
the other hand, we have $i<b_2$ because $i< b = \max(b_1, b_2)$.
It hence follows that $i\in [a_2, b_2)$ if $i\notin [a_1, b_1)$.
No matter in which case, i.e., either $i\in [a_1, b_1)$ or $i\in
[a_2, b_2)$, we can have $ W_1^{'}(\mathcal{C}) \subseteq
W_1(\mathcal{C}) \cup W_2(\mathcal{C})$. Thus, $W_1(\mathcal{C})
\cup W_2(\mathcal{C}) = W_1^{'}(\mathcal{C})$. Consequently, we
can obtain a smaller-sized partition of $W(\mathcal{C})$ by
replacing two sets $W_1(\mathcal{C})$ and $W_2(\mathcal{C})$ of
the current partition $\mathbb{W}(\mathcal{C})$ with one set
$W_1^{'}(\mathcal{C})$, which however contradicts the fact that
$\mathbb{W}(\mathcal{C})$ attains the minimum cardinality. $\qed$

\begin{lem} \label{lem:108}
Let $\mathcal{C}$ be a (not necessarily simple) cycle with
$W_1(\mathcal{C}) = \{ i\cdot (i+1) \ | \ a\leq i < b\}$ being an
element of $\mathbb{W}(\mathcal{C})$. If there exists a vertex
$c\in \Sigma (\mathcal{C})$ such that $c\notin [a,b]$, then
$\mathcal{C}$ is a conflict-cycle.
\end{lem}

\noindent {\em Proof.} \indent We first assume that $c< a$. Define
$a^+ = \{ i|i\ge a\} \cup \{ i\cdot (i+1) | i\ge a\}$ and $a^- =
\{ i|i< a\} \cup \{ i\cdot (i+1) | i<a\}$. Then, $a^+ \cup
 a^-$ is a partition of $V$. Note that there exists
in $F$ exactly one arc from $a^+$ to $a^-$ and exactly one arc
from $a^-$ to $a^+$, i.e., $a \to (a-1) _F\cdot a$ and $(a-1)
_F\cdot a \to a$, respectively. Suppose that $\mathcal{C}$ does
not contain any arc from $X$. Since $\mathcal{C}$ contains
vertices in both $a^+ $ and $a^-$ (resp. $b$ and $c$), it thus
contains an arc $u\to v$ with $u \in a^+$ and $v\in a^-$. We must
have $u\to v \in F$; otherwise, $u\to v \in D$ implies that $u\to
v \in X$ since $u>v$. Consequently, we can only have $u=a$ and $v=
(a-1) \cdot a$ by the definitions of $a^+$ and $a^-$. So,
$\mathcal{C}$ uses the vertex $(a-1) \cdot a$. However,
$W_1(\mathcal{C}) = \{ i\cdot (i+1) \ | \ a\leq i < b\}$ is an
element of $\mathbb{W}(\mathcal{C})$, which, by definition,
implies that $\mathcal{C}$ does not use the vertex $(a-1) \cdot
a$; a contradiction. Therefore, $\mathcal{C}$ must contain an arc
from $X$. Now we suppose that $\mathcal{C}$ does not contain any
arc from $Y$. Once again, since $\mathcal{C}$ contains vertices in
both $a^+ $ and $a^-$, it thus contains an arc $u\to v$ with $u
\in a^-$ and $v\in a^+$. We must have $u\to v \in F$; otherwise,
$u\to v \in D$ implies that $u\to v \in Y$ since $u<v$.
Consequently, we can only have $u=(a-1) \cdot a$ and $v= a$. So,
$\mathcal{C}$ also necessarily uses the vertex $(a-1) \cdot a$. As
we show above, it would lead to a contradiction. Therefore,
$\mathcal{C}$ must contain an arc from $Y$ too. It turns out that
$\mathcal{C}$ is a conflict-cycle.

In case of $c>b$, we may define $b^+ = \{ i|i> b\} \cup \{ i\cdot
(i+1) | i\ge b\}$ and $b^- = \{ i|i\le b\} \cup \{ i\cdot (i+1) |
i<b\}$. Then, by using the same arguments as above, we can also
show that $\mathcal{C}$ is a conflict-cycle. $\qed$ \\

\begin{lem} \label{lem:110}
Let $\pi$ be a total order that contains every adjacency in the
set $ \{ i\cdot (i+1) \ | \ a\leq i < b\}$. Then, either the
sequence $a\ (a+1)\ (a+2)\ \cdots \ b$ or $b\ (b-1)\ (b-2)\ \cdots
\ a$ is an interval of $\pi$.
\end{lem}

\noindent {\em Proof.} \indent Recall that an adjacency $i\cdot
(i+1)$ implies the occurrence of an interval either $i\ (i+1)$ or
$(i+1) \ i$, but not both, in $\pi$. We first consider the
adjacency $a\cdot (a+1)$, for which the interval either $a\ (a+1)$
or $(a+1) \ a$ would occur in $\pi$. We distinguish these two
cases when the next adjacency $(a+1)\cdot (a+2)$ is considered. In
the first case of the interval $a\ (a+1)$, in order to obtain the
adjacency $(a+1)\cdot (a+2)$ in $\pi$, the element $(a+2)$ can
only appear immediately after the element $(a+1)$, resulting in
the interval $a\ (a+1)\ (a+2)$. In the second case of the interval
$(a+1) \ a$, in order to obtain the adjacency $(a+1)\cdot (a+2)$
in $\pi$, the element $(a+2)$ can only appear immediately before
the element $(a+1)$, resulting in the interval $(a+2)\ (a+1)\ a$.
Continue this process with the remaining adjacencies in the
increasing order of elements. It would necessarily end up with an
interval either $a\ (a+1)\ (a+2)\ \cdots \ b$ or $b\ (b-1)\ (b-2)\
\cdots \ a$ in $\pi$. $\qed$

\begin{lem} \label{lem:112}
Let $\pi$ be a total order that contains every adjacency in the
set $ \{ i\cdot (i+1) \ | \ a\leq i < b\}$. Assume that there
exists in $G_\Pi$ an arc $i_1 \to i_2\in D$, where $a\leq i_1 \leq
b$ and $a\leq i_2 \leq b$. If $i_1 < i_2$ (resp., $i_1 > i_2$),
then the sequence $a\ (a+1)\ (a+2)\ \cdots b$ (resp., $b\ (b-1)\
(b-2)\ \cdots a$) is an interval of $\pi$.
\end{lem}

\noindent {\em Proof.} \indent The proof is given only for the
case of $i_1 < i_2$. We know from Lemma~\ref{lem:110} that $\pi$
contains either the interval $a\ (a+1)\ (a+2)\ \cdots\ i_1 \cdots\
i_2 \cdots \ b$ or $b\ (b-1)\ (b-2)\ \cdots\ i_2 \cdots\ i_1
\cdots \ a$. On the other hand, we have $i_1 \prec_\pi i_2$, since
there exists an arc $i_1 \to i_2\in D$. Consequently, the interval
$b\ (b-1)\ (b-2)\ \cdots\ i_2 \cdots\ i_1 \cdots \ a$ could not
appear in $\pi$. $\qed$ \\

We wish to distinguish two types of conflict-cycles.  A
conflict-cycle $\mathcal{C}$ is said to be of type I if there
exist two vertices $a$ and $b$ in $\Sigma(\mathcal{C})$ such that
$V(\mathcal{C}) = \{ i\cdot (i+1) \ | \ a\leq i < b\} \cup \{ i \
| \ a\leq i \leq b\}$; otherwise, it is said to be of type II. For
example, in the adjacency-order graph shown in
Figure~\ref{fig:An-adjacency-graph}, the cycle $1\to 2\to 2\cdot 3
\to 3\to 3\cdot 4 \to 4 \to 4\cdot 5\to 5 \to 3\to 2\cdot 3 \to
2\to 1\cdot 2\to 1$ is a conflict-cycle of type I, while both
$2\to 5\to 3\to 2\cdot 3\to 2$ and  $2\to 4\to 4\cdot 5 \to 5\to
3\to 2\cdot 3\to 2$ are conflict-cycles of type II.
Lemmas~\ref{lem:113} and~\ref{lem:114} below follows from the
above definitions in a straightforward way.

\begin{lem} \label{lem:113}
Let $\mathcal{C}$ is a (not necessarily simple) conflict-cycle of
type 1. Then, $|\mathbb{W} (\mathcal{C})| = 1$.
\end{lem}

\begin{lem} \label{lem:114}
Let $\mathcal{C}$ is a (not necessarily simple) cycle with
$W_1(\mathcal{C}) = \{ i\cdot (i+1) \ | \ a\leq i < b\}$ being an
element of $\mathbb{W}(\mathcal{C})$. Then, $\mathcal{C}$ is a
conflict-cycle of type II iff there exists a vertex $c\in \Sigma
(\mathcal{C})$ such that $c\notin [a,b]$.
\end{lem}

By considering Lemmas~\ref{lem:106} and~\ref{lem:108}, we can
further obtain the following lemma.

\begin{lem} \label{lem:115}
Let $\mathcal{C}$ be a (not necessarily simple) cycle with
$|\mathbb{W} (\mathcal{C}) | \ge 2$. Then, $\mathcal{C}$ is a
conflict-cycle of type II.
\end{lem}

The first implication of our new definition of conflict-cycle is
that a conflict-cycle does not necessarily contain a simple
conflict-subcycle.

\begin{lem} \label{lem:116}
If $\mathcal{C}$ is a conflict-cycle of type I, then it cannot be
a simple cycle.
\end{lem}

\noindent {\em Proof.} \indent By contradiction, suppose that
$\mathcal{C}$ is simple. By definition of a type I conflict-cycle,
there exist two vertices $a$ and $b$ such that $V(\mathcal{C}) =
\{ i\cdot (i+1) \ | \ a\leq i < b\} \cup \{ i \ | \ a\leq i \leq
b\}$. Since $\mathcal{C}$ is simple, every vertex in
$V(\mathcal{C})$ is adjacent to exactly two distinct vertices in
$\mathcal{C}$; therefore, every vertex has indegree and outdegree
both exactly one in $\mathcal{C}$. Knowing that every vertex
$i\cdot (i+1) \in W$ has only two distinct adjacent vertices in
$G_\Pi$, i.e., $i$ and $(i+1)$, we can deduce that, for every
vertex $i$ such that $a<i<b$, it is adjacent to both $(i-1)\cdot
i$ and $i\cdot (i+1)$ by using arcs from $F$. And, the vertex $a$
is adjacent to $a\cdot (a+1)$ and the vertex $b$ is adjacent to
$(b-1)\cdot b$, both using arcs also from $F$. Consequently,
$\mathcal{C}$ shall contain an arc between $a$ and $b$ so that
both vertices have degree two (because any other vertices can no
longer be incident to an arc of $D(\mathcal{C})$). Moreover, this
arc is the only arc that $\mathcal{C}$ has from $D(\mathcal{C})$,
which contradicts the fact that a conflict-cycle shall contain at
least two arcs from $D(\mathcal{C})$, i.e., at least one from
$X(\mathcal{C})$ and at least one from $Y(\mathcal{C})$. $\qed$

\begin{lem} \label{lem:118}
If $\mathcal{C}$ is a non-simple conflict-cycle of type II, then
it must contain a simple conflict-subcycle of type II.
\end{lem}

\noindent {\em Proof.} \indent Let $\mathbb{W}(\mathcal{C}) = \{
W_1(\mathcal{C}), W_2(\mathcal{C}), \cdots, W_l(\mathcal{C}) \}$.
Since $\mathcal{C}$ is not simple, there exists a vertex $u$ used
twice in it such that $\mathcal{C} =u \xrightarrow{P}{}^+ u
\xrightarrow{Q}{}^+ u$. We can further assume that $u\in
\Sigma(\mathcal{C})$. If initially we have $u\in W(\mathcal{C})$
such that $u=a\cdot (a+1)$, then $\mathcal{C}$ uses both vertices
$a$ and $(a+1)$ at least twice because it uses the vertex
$u=a\cdot (a+1)$ twice. So, we may substitute $u$ by $a$ to write
$\mathcal{C} =u \xrightarrow{P}{}^+ u \xrightarrow{Q}{}^+ u$.

Let $\mathcal{C}_1 =u \xrightarrow{P}{}^+ u$ and $\mathcal{C}_2 =
u \xrightarrow{Q}{}^+ u$. Apparently, $\mathcal{C}_1$ and
$\mathcal{C}_2$ are two subcycles of $\mathcal{C}$, so we write
$\mathbb{W}(\mathcal{C}_1) = \{ W_1(\mathcal{C}_1),
W_2(\mathcal{C}_1), \cdots, W_{l_1}(\mathcal{C}_1) \}$ and
$\mathbb{W}(\mathcal{C}_2) = \{ W_1(\mathcal{C}_2),
W_2(\mathcal{C}_2), \cdots, W_{l_2}(\mathcal{C}_2) \}$, where
$l_1\ge 1$ and $l_2\ge 1$. Note that every element of
$\mathbb{W}(\mathcal{C}_1)$ and of $\mathbb{W}(\mathcal{C}_2)$ is
a subset of an element of $\mathbb{W}(\mathcal{C})$. Below we
distinguish two possible cases.

In the first case, we assume that there exist an element of
$\mathbb{W}(\mathcal{C}_1)$ and an element of
$\mathbb{W}(\mathcal{C}_2)$ (say, $W_1(\mathcal{C}_1) = \{ i\cdot
(i+1) \ | \ a_{11}\leq i < b_{11}\}$ and $W_1(\mathcal{C}_2) = \{
i\cdot (i+1) \ | \ a_{21}\leq i < b_{21}\}$, respectively) such
that both are the subsets of a same element of
$\mathbb{W}(\mathcal{C})$ (say, $W_1(\mathcal{C}) = \{ i\cdot
(i+1) \ | \ a_1\leq i < b_1 \}$). It hence implies that $a_1\leq
a_{11}< b_{11}\leq b_1$ and $a_1\leq a_{21}< b_{21}\leq b_1$.
Since $\mathcal{C}$ is a conflict-cycle of type II, by
Lemma~\ref{lem:114}, there exists a vertex $c_1\in \Sigma
(\mathcal{C})$ such that $c_1\notin [a_1,b_1]$. Thus, we have both
$c_1\notin [a_{11},b_{11}]$ and $c_1\notin [a_{21},b_{21}]$. Note
that the vertex $c_1$ appears on the cycle either $\mathcal{C}_1$
or $\mathcal{C}_2$. If $c_1$ appears on $\mathcal{C}_1$, then
$\mathcal{C}_1$ is a conflict-cycle (by Lemma~\ref{lem:108}).
Otherwise, $c_2$ must appear on $\mathcal{C}_2$. By
Lemma~\ref{lem:108} once again, $\mathcal{C}_2$ would be a
conflict-cycle. Moreover, this conflict-cycle, no matter
$\mathcal{C}_1$ or $\mathcal{C}_2$, is of type II (by
Lemma~\ref{lem:115}).

In the second case, we assume that no two elements of
$\mathbb{W}(\mathcal{C}_1)$ and $\mathbb{W}(\mathcal{C}_2)$ are
the subsets of a same element of $\mathbb{W}(\mathcal{C})$.
Consider the first elements of $\mathbb{W}(\mathcal{C}_1)$ and
$\mathbb{W}(\mathcal{C}_2)$, and write them as $W_1(\mathcal{C}_1)
= \{ i\cdot (i+1) \ | \ a_{11}\leq i < b_{11}\}$ and
$W_1(\mathcal{C}_2) = \{ i\cdot (i+1) \ | \ a_{21}\leq i <
b_{21}\}$, respectively. Note that $W_1(\mathcal{C}_1)$ and
$W_1(\mathcal{C}_2)$ are the subsets of two distinct elements
(say, $W_1(\mathcal{C}) = \{ i\cdot (i+1) \ | \ a_1\leq i < b_1
\}$ and $W_2(\mathcal{C}) = \{ i\cdot (i+1) \ | \ a_2\leq i < b_2
\}$) of $\mathbb{W}(\mathcal{C})$, respectively). Thus, we have
$[a_{11}, b_{11}] \subseteq [a_1, b_1]$ and $[a_{21}, b_{21}]
\subseteq [a_2, b_2]$ and, furthermore, $[a_{11}, b_{11}] \cap
[a_{21}, b_{21}] = \emptyset$ since $[a_1, b_1] \cap [a_2, b_2] =
\emptyset$. It then follows that we have either $u\notin [a_{11},
b_{11}]$ or $u\notin [a_{21}, b_{21}]$. If $u\notin [a_{11},
b_{11}]$, $\mathcal{C}_1$ would be a conflict-cycle of type II. If
$u\notin [a_{21}, b_{21}]$, $\mathcal{C}_2$ would be a
conflict-cycle of type II.

In either case above, we already show that there exists a
conflict-subcycle of type II for $\mathcal{C}$. If this
conflict-subcycle is not simple, we may apply the above process
recursively, which necessarily ends up with a simple
conflict-subcycle of type II. $\qed$ \\

Although the following theorem appears as a verbatim account of
Theorem 4 in \cite{BF10}, they are literally not the same because
conflict-cycles are defined in different ways. Consequently, the
corresponding proof given in~\cite{BF10} is not sufficient.

\begin{thm} \label{thm:conflict-cycle}
Let $\Pi$ be a partial order, $G_\Pi = (V, E)$ its adjacency-order
graph (with $V=\Sigma \cup W$ and $E= D\cup F$), and
$W^{'}\subseteq W$. Then there exists a total order $\pi$ over
$\Sigma$, compatible with $\Pi$, and containing every adjacency
from $W^{'}$ iff $G_\Pi [W^{'}\cup \Sigma]$ has no conflict-cycle.
\end{thm}

\noindent {\em Proof.} \indent ($\Rightarrow$) Let $\pi$ be a
linearization of $\Pi$ containing every adjacency of $W^{'}$. We
suppose, by contradiction, that there exists in $G_\Pi [W^{'} \cup
\Sigma]$ a conflict-cycle $\mathcal{C}$. Below we distinguish two
cases, depending on whether $\mathcal{C}$ is of type I or of type
II.

In the first case, $\mathcal{C}$ is assumed to be of type I. By
definition, there exist two integers $a$ and $b$ such that
$W(\mathcal{C}) = \{ i\cdot (i+1) \ | \ a\leq i < b\}$ and
$\Sigma(\mathcal{C}) = \{ i \ | \ a\leq i \leq b\}$. Since
$\mathcal{C}$ is a conflict-cycle, there exists an arc $i_1 \to
j_1 \in X$ such that $a\leq j_1 < i_1 \leq b$ and an arc $i_2 \to
j_2 \in Y$ such that $a\leq i_2 < j_2 \leq b$. By
Lemma~\ref{lem:112}, the arc $i_1 \to j_1$ implies that the
sequence $b\ (b-1)\ (b-2)\ \cdots a$ appears as an interval of
$\pi$, while at the same time the arc $i_2 \to j_2$ implies that
the sequence $a\ (a+1)\ (a+2)\ \cdots b$ appears as an interval of
$\pi$; a contradiction.

In the second case, $\mathcal{C}$ is assumed to be a
conflict-cycle of type II. W.l.o.g, we may further assume that
$\mathcal{C}$ is a simple conflict-cycle of type II (by
Lemma~\ref{lem:118}). Let $\mathcal{C} = v_0 \to v_1 \to \cdots
\to v_l = v_0$ where all the vertices are pairwise distinct except
$v_0 = v_l$. Let $i_0=0,\ i_1,\dots, i_{h-1}, i_h =l$ be the
increasing sequence of indices such that $v_{i_j} \to v_{i_j+1}
\in D$ for all $j$ such that $0\leq j < h$. Note that $h\ge 2$
(because $|D(\mathcal{C})|\ge 2$) and, for all $j$, we have
$v_{i_j}\in \Sigma$. Let us prove that for all $j<h$, the ordering
relation $v_{i_j} \prec_\pi v_{i_{j+1}}$ holds. The case where
$i_{j+1} = i_j+1$ is easy, since the arc $v_{i_j} \to v_{i_j+1}
\in D$ implies that $v_{i_j} \prec_\Pi v_{i_{j+1}}$ (by
construction of $G_\Pi$) and $v_{i_j} \prec_\pi v_{i_{j+1}}$
(since $\pi$ is compatible with $\Pi$). Now, assume there are
several arcs between $v_{i_j}$ and $v_{i_{j+1}}$, i.e.,
$v_{i_{j+1}} = v_{i_j} +m$ with $m\ge 2$. Let $P=v_{i_j+1} \to
v_{i_j+2} \to \cdots \to v_{i_j+m}$, in which all the arcs are
from $F$ and $v_{i_j+1}, v_{i_j+m}\in \Sigma$. If $v_{i_j+1} <
v_{i_j+m}$, then $W(P)=\{i\cdot (i+1)\ |\ v_{i_j+1}\leq i <
v_{i_j+m} \}$ and $\Sigma(P)=\{i\ |\ v_{i_j+1}\leq i \leq
v_{i_j+m} \}$. By Lemma~\ref{lem:110}, the sequence $v_{i_j+1} \
(v_{i_j+1}+1) \ (v_{i_j+1}+2) \cdots \ v_{i_j+m}$ appears as an
interval of $\pi$. If $v_{i_j+1} > v_{i_j+m}$, then $W(P)=\{i\cdot
(i+1)\ |\ v_{i_j+m}\leq i < v_{i_j+1} \}$ and $\Sigma(P)=\{i\ |\
v_{i_j+m}\leq i \leq v_{i_j+1} \}$. Again, by Lemma~\ref{lem:110},
the sequence $v_{i_j+m} \ (v_{i_j+m}-1) \ (v_{i_j+m}-2) \cdots \
v_{i_j+1}$ appears as an interval of $\pi$. In either case, all
the vertices in $\Sigma(P)$ therefore appear as an interval of
$\pi$. Note that $v_{i_j}$ is a vertex distinct from $v_{i_{j+1}}$
(since $h\ge 2$), and from other vertices in the set $\Sigma(P)$
as well (since each of them is the source of an arc from $F$ in
$\mathcal{C}$, where $v_{i_{j+1}}$ is the source of an arc from
$D$ in $\mathcal{C}$). Consequently, $v_{i_j}$ cannot appear
inside either of the intervals $v_{i_j+1} \ (v_{i_j+1}+1) \
(v_{i_j+1}+2) \cdots \ v_{i_j+m}$ or $v_{i_j+m} \ (v_{i_j+m}-1) \
(v_{i_j+m}-2) \cdots \ v_{i_j+1}$ of $\pi$. As $v_{i_j}$ precedes
$v_{i_j+1}$ in $\Pi$ (and thus in $\pi$), we have
$v_{i_j}\prec_\pi v_{i}$ for all $i\in [i_j+1, i_j+m]$, and
particularly, $v_{i_j}\prec_\pi v_{i_{j+1}}$.

In conclusion, we have $v_{i_j}\prec_\pi v_{i_{j+1}}$ for all
$j<h$ and $v_{i_h} = v_{i_0}$, leading to a contradiction since
there is no cycle in the ordering relation $\prec_\pi$. Therefore,
the subgraph $G_\Pi [W^{'}\cup \Sigma]$ does not contain any
conflict-cycle.

($\Leftarrow$) ({\em constructive proof}) We use the following
method to construct a linearization $\pi$ of $\Pi$ containing all
adjacencies of $W^{'}$, where the subgraph $G^{'} = G_\Pi [W^{'}
\cup \Sigma]$, is assumed to contain no conflict-cycles. We denote
by $V_1,\dots, V_k$ the strongly connected components of $G^{'}$,
ordered by topological order (i.e., if $u,v\in V_i$, there exists
a path from $u$ to $v$; moveover, if $u\in V_i$ and $v\in V_j$ and
there exists a path $u\to^* v$ in $G^{'}$, then $i\leq j$). Then,
we sort the elements of each set $V_i \cap \Sigma$ in descending
order of integers if there exists an arc from $X$ connecting two
vertices in $V_i \cap \Sigma$; otherwise, sort them in ascending
order. The resulting sequence is denoted by $\mu_i$, and the
concatenation $\mu_1\cdot \mu_2 \cdot \dots$ gives $\pi$, a total
order over $\Sigma$. We now check that $\pi$ contains every
adjacency in $W^{'}$ and is compatible with $\Pi$.

Let $a\cdot (a+1) \in W^{'}$. Vertices $a$ and $a+1$ are in the
same strong connected component $V_i$, due to the arcs $a
\leftrightarrow a\cdot (a+1) \leftrightarrow (a+1)$. Those two
elements are obviously consecutive in the corresponding $\mu_i$,
and appear as an adjacency in $\pi$.

To show that $\pi$ is compatible with $\Pi$, it suffices by
showing that $a\prec_\pi b$ holds for every arc $a\to b \in D$. By
contradiction, suppose that there exist two distinct elements
$a,b\in \Sigma$ such that $a\to b \in D$ but $b\prec_\pi a$. We
denote by $i$ and $j$ the indices such that $a\in V_i$ and $b\in
V_j$. Since $b\prec_\pi a$, we have $j\leq i$, and since $a\to b
\in D$ (the arc $a\to b$ in $G^{'}$ as well), we have $i\leq j$.
We thus deduce that $i=j$; therefore, $a$ and $b$ share the same
strong connected component. If $a\to b \in X$, then $a>b$ and
$a\prec_\pi b$ (by the construction of $\pi$); a contradiction.
Therefore, $a\to b \in Y$, which then implies that $a < b$. Since
$b\prec_\pi a$, by the construction of $\pi$ once again, there
must exist an arc $c\to d\in X$ such that $c$ and $d$ belong to
the same strong connected component as $a$ and $b$. It hence
follows that there exists a path $P_1$ from $b$ to $c$ in $G^{'}$
and also a path $P_2$ from $d$ to $a$ in $G^{'}$. Consequently, we
obtain a cycle $a\to_Y b \xrightarrow{P_1}{}^* c\to_X d
\xrightarrow{P_2}{}^* a$, which, by definition, is a
conflict-cycle in $G^{'}$; a contradiction. $\qed$

\section{Approximation} \label{sec:approximation}

\subsection{Approximation of the \textsc{MBL} problem} \label{sec:MBL}

To assist in solving the minimum breakpoint linearization problem,
the above theorem motivates us to formulate a new combinatorial
optimization problem on an adjacency-order graph. Given an
adjacency-order graph $G_\Pi=(V,E)$, where $V=\Sigma \cup W$ with
$E=D\cup F$ and $D=X\cup Y$, a subset $W^{''}$ of $W$ is called a
{\em breakpoint vertex set} if the deletion of vertices in
$W^{''}$ leaves the induced subgraph $G_\Pi[V-W^{''}]$ without any
cycle using arcs from both $X$ and $Y$. The {\em minimum
breakpoint vertex set} (MBVS) problem is thus defined as the
problem of finding a breakpoint vertex set with minimum
cardinality. Theorem~\ref{thm:conflict-cycle} leads to the
following corollary.
\begin{cor}
The value $k$ of an optimal solution of \textsc{MBL}($\Pi$) is the
size of the minimum breakpoint vertex set of $G_\Pi$.
\end{cor}
It implies that an approximation algorithm for the MBVS problem
can be translated into an approximation algorithm for the MBL
problem with the same ratio.

As in \cite{BF10}, we denote by SCC-sort$()$ an algorithm that
decomposes a directed graph into its strong connected components
and then topologically sorts these components. Also, let sort$()$
denote an algorithm that sorts the integer elements in each
strongly connected component either in a descending order or an
ascending order, as we described in the constructive proof of
Theorem~\ref{thm:conflict-cycle}. Note that a different definition
of sort$()$ was used in \cite{BF10}, which always sorts integers
in an ascending order. Table~\ref{alg:MBL} summarizes the
algorithm that is used to approximate the MBL problem,
\textsc{Approx-MBL}. It is derived from the constructive proof of
Theorem~\ref{thm:conflict-cycle}, and relies on an approximation
algorithm for the MBVS problem that we are going to describe in
the next subsection. Its correctness follows from
Theorem~\ref{thm:conflict-cycle}.

\begin{table}[t] \centering
\begin{tabular}{|l|}
  \hline
   \rowcolor[rgb]{0.9,0.9,0.9} {\bf Algorithm} \textsc{Approx-MBL}  \\
  \hline
  {\bf input} A directed acyclic graph $\Pi = (\Sigma, D)$ \\
  {\bf output} A linearization $\pi$ of $\Pi$ \\
  {\bf begin}  \\
  \ \ \ Create the adjacency-order graph $G_\Pi =(V,E)$ of $\Pi$; \\
  \ \ \ $W^{''} \leftarrow \textsc{Approx-MBVS}(G_\Pi)$;   \\
  \ \ \ $W^{'} \leftarrow W-W^{''}$; \\
  \ \ \ $(V_1, V_2, \dots, V_h) \leftarrow$ SCC-sort$(G_\Pi[W^{'} \cup \Sigma])$; \\
  \ \ \ {\bf for} $i \leftarrow 1$ {\bf to} $h$   \\
  \ \ \ \quad $\mu_i \leftarrow$ sort$(V_i\cap \Sigma)$; \\
  \ \ \ $\pi \leftarrow \mu_1 \cdot \mu_2 \cdots \mu_h$; \\
  \ \ \ {\bf return} $\pi$; \\
  {\bf end} \\
  \hline
\end{tabular} \caption{An $(m^2 + 2m -1)$-approximation for the \textsc{MBL}
problem.}  \label{alg:MBL}
\end{table}

\subsection{Approximation of the \textsc{MBVS} problem} \label{sec:MBVS}

We start this subsection by introducing several more definitions.
As similarly defined in \cite{BF10}, a path $u
\xrightarrow{R}{}_D^* v$ in $(\Sigma, D)$ is said to be a {\em
shortcut} of a type II conflict-cycle $\mathcal{C}$, if:
\begin{enumerate}
 \item[-] $u,v\in \Sigma(\mathcal{C})$ (we write $P$ and $Q$ the
 paths such that $\mathcal{C} = v \xrightarrow{P}{}^+ u \xrightarrow{Q}{}^+
 v$),
 \item[-] the cycle $\mathcal{C}^{'} = v \xrightarrow{P}{}^+ u \xrightarrow{R}{}_D^* v$ is also a conflict-cycle of type II,
 \item[-] $W(Q)\ne \emptyset$ (using the shortcut removes at least one adjacency).
\end{enumerate}
A type II conflict-cycle is said to be {\em minimal} if it has no
shortcut. On the other hand, a type I conflict-cycle is said to be
{\em minimal} if there does not exist another type I
conflict-cycle with a strict subset of $W(\mathcal{C})$. Note that
the definition of shortcut does not apply to the conflict-cycles
of type I. The following lemma ensures that removing minimal
conflict-cycles is enough to remove all the conflict-cycles.

\begin{lem} \label{lem:204}
If an adjacency-order graph contains a conflict-cycle, then it
also contains a minimal conflict-cycle.
\end{lem}

\noindent {\em Proof.} \indent Let $\mathcal{C}$ be a
conflict-cycle. Suppose that $\mathcal{C}$ is not minimal. If it
is a conflict-cycle of type I, by definition, we may find another
type I conflict-cycle $\mathcal{C}^{'}$ with $|W(\mathcal{C}^{'})|
< |W(\mathcal{C})|$; if it is a conflict-cycle of type II, we may
use the shortcut to create another conflict-cycle
$\mathcal{C}^{'}$ of type I also having $|W(\mathcal{C}^{'})| <
|W(\mathcal{C})|$. Applied recursively, this process necessarily
ends with a minimal conflict-cycle. $\qed$

\begin{lem} \label{lem:306}
Let $\mathcal{C}$ be a minimal conflict-cycle. Then, $\mathcal{C}$
is simple if and only if it is of type II.
\end{lem}

\noindent {\em Proof.} \indent ($\Rightarrow$) Since $\mathcal{C}$
is a simple conflict-cycle, by Lemma~\ref{lem:116}, $\mathcal{C}$
cannot be of type I. Therefore, $\mathcal{C}$ must be a
conflict-cycle of type II.

($\Leftarrow$) By contradiction, suppose that $\mathcal{C}$ is not
simple. Since $\mathcal{C}$ is of type II, by Lemma~\ref{lem:118},
it must contain a simple conflict-subcycle $\mathcal{C}^{'}$ of
type II. So, we may write $\mathcal{C} = u
\xrightarrow{\mathcal{C}^{'}}{}^+  u \xrightarrow{Q}{}^+ u$, where
$u\in \Sigma(\mathcal{C})$ (see the proof of Lemma~\ref{lem:118}).
Let $R= u\to_\emptyset u$ be a path with an empty arc set. We can
see that $\mathcal{C}^{'} = u \xrightarrow{\mathcal{C}^{'}}{}^+  u
\xrightarrow{R}{}^* u$ is a conflict-cycle and that $W(Q) \ne
\emptyset$ (since $Q$ is a cycle of $\mathcal{C}$), so the path
$R$ is a shortcut of $\mathcal{C}$. It hence leads to a
contradiction that $\mathcal{C}$ is indeed given as a minimal
conflict-cycle. $\qed$ \\

Let $\mathcal{C}$ be a cycle in $G_\Pi$ with
$\mathbb{W}(\mathcal{C}) = \{ W_1(\mathcal{C}), W_2(\mathcal{C}),
\cdots, W_l(\mathcal{C}) \}$, where $W_h(\mathcal{C}) = \{ i\cdot
(i+1) \ | \ a_h\leq i < b_h\}$, for each $1\le h\le l$. We call
the vertices $a_h$ and $b_h$ the {\em joints} of $\mathcal{C}$
and, in particular, $a_h$ the {\em low joint}. Given a vertex
$i\cdot (i+1)\in W(\mathcal{C})$, we say that $a_h$ and $b_h$ are
the two joints {\em associated} to $w$ in $\mathcal{C}$ if $a_h\le
i < b_h$. Note that joints are also defined in \cite{BF10}, but
not in the same way.

Our approximation algorithm for the MBVS problem is summarized in
Table~\ref{alg:MBVS}. As we can see, it consists of two main
phases. In the first phrase, the adjacency-order graph $G_\Pi$ is
repeatedly induced by deleting a set of low joints of a minimal
type II conflict-cycle until there are no more minimal type II
conflict-cycles (except for one case where $m = 1$ and
$|\mathbb{W}(\mathcal{C})| =1$). In the second phase, the
previously induced subgraph is further repeatedly induced by
deleting the only two joints of a type I conflict-cycle until
there are no more minimal type I conflict-cycles. It is worth
noting that finding a minimal type II conflict-cycle is quite
challenging, due to the presence of type I conflict-cycles in the
adjacency-order graph. We will discuss the polynomial-time
algorithms for finding type I and type II conflict-cycles in
Subsection~\ref{sec:time}.

\begin{table}[t] \centering
\begin{tabular}{|l|}
  \hline
   \rowcolor[rgb]{0.9,0.9,0.9} {\bf Algorithm} \textsc{Approx-MBVS}  \\
  \hline
   {\bf input} An adjacency-order graph $G_\Pi (V, E)$ \\
   {\bf output} A breakpoint vertex set $W^{''}$ \\
   {\bf begin}  \\
  \ \ \ $W^{''} \leftarrow \emptyset$; \\
  \ \ \ {\bf while} there exists in $G_\Pi [V- W^{''}]$ a minimal type II conflict-cycle $\mathcal{C}$  \\
  \ \ \ \quad {\bf if} $m = 1$ and $|\mathbb{W}(\mathcal{C})| =1$ \\
  \ \ \ \quad \quad $J \leftarrow$ the set of joints of $\mathcal{C}$; \\
  \ \ \ \quad {\bf else}  \\
  \ \ \ \quad \quad $J \leftarrow$ the set of low joints of $\mathcal{C}$; \\
  \ \ \ \quad $W^{''} \leftarrow W^{''} \cup \{ e^F\ :\ e\in J \}$; \\
  \ \ \ {\bf while} there exists in $G_\Pi [V- W^{''}]$ a minimal type I conflict-cycle $\mathcal{C}$   \\
  \ \ \ \quad $J \leftarrow$ the set of joints of $\mathcal{C}$; \\
  \ \ \ \quad $W^{''} \leftarrow W^{''} \cup \{ e^F\ :\ e\in J \}$; \\
  \ \ \ {\bf return} $W^{''}$; \\
   {\bf end}  \\
  \hline
\end{tabular} \caption{An $(m^2 + 2m -1)$-approximation for the \textsc{MBVS}
problem} \label{alg:MBVS}
\end{table}

\section{Performance Analysis} \label{sec:performance}

\subsection{Approximation ratio}

If $\mathcal{C}$ is given as a minimal conflict-cycle of type II,
it must be simple by Lemma~\ref{lem:306}. Hence, a joint $e$ of
$\mathcal{C}$ has exactly two incident arcs, one belonging to
$D(\mathcal{C})$ and the other belonging to $F(\mathcal{C})$. In
this case, we denote by $e^F$ the other vertex (rather than $e$)
of the arc from $F(\mathcal{C})$, and by $e^D$ the other vertex
(rather than $e$) of the arc from $D(\mathcal{C})$.

As defined in \cite{BF10}, for each $u\in \Sigma$, we denote $I(u)
\subseteq \{1,\dots, m\}$ the number of the genetic maps in which
$u$ appears. Clearly, $I(u)\ne \emptyset$. For each arc $u\to_D v
\in D$, we use $\eta(u\to_D v)$ to denote the numbering of a
genetic map in which $u$ and $v$ appear in consecutive blocks. So,
$\eta(u\to_D v) \in I(u) \cap I(v)$. Given a minimal type II
conflict-cycle $\mathcal{C}$, we extend the notation $\eta$ to
each of its joints $e$: let $\eta(e) = \eta(e^D\to e)$ if
$\mathcal{C}$ uses the arc $e^D\to e$; otherwise, let $\eta(e) =
\eta(e\to e^D)$.

\begin{lem} \cite{BF10} \label{lem:400}
Let $e\to f$ be an arc of $D$, and let $u\in \Sigma$ such that
$\eta (e\to_D f) \in I(u)$. Then one of the paths $e\to^* u$ or
$u\to^* f$ appears in the graph $(\Sigma, D)$.
\end{lem}

\begin{lem} \cite{BF10} \label{lem:401a}
Let $\mathcal{C}$ be a (not necessarily simple) cycle of $G_\Pi$.
Let $c\in \Sigma$, such that there exists $a,b\in
\Sigma(\mathcal{C})$ with $a\le c < b$. Then, one of the following
propositions is true:
\begin{enumerate}
 \item[(i)] $\mathcal{C}$ contains an arc $u\to_X v$ with $v\le c
 < u$;
 \item[(ii)] $\mathcal{C}$ contains both arcs $c+1\to_F c\cdot
 (c+1)$ and $c\cdot (c+1)\to_F c$.
\end{enumerate}
\end{lem}

We can further obtain the following lemma, which can be proved by
using the same arguments as those for proving the preceding lemma.

\begin{lem} \label{lem:401b}
Let $\mathcal{C}$ be a (not necessarily simple) cycle of $G_\Pi$.
Let $c\in \Sigma$, such that there exists $a,b\in
\Sigma(\mathcal{C})$ with $a \le c< b$. Then, one of the following
propositions is true:
\begin{enumerate}
 \item[(i)] $\mathcal{C}$ contains an arc $u\to_Y v$ with $u \le c< v$;
 \item[(ii)] $\mathcal{C}$ contains both arcs $c\to_F c\cdot
 (c+1)$ and $c\cdot (c+1)\to_F c+1$.
\end{enumerate}
\end{lem}

\noindent {\em Proof.} \indent Define $c^{+} = \{ d| d>c\} \cup \{
d\cdot (d+1) | d>c\}$ and $c^- = \{ d| d\le c\} \cup \{ d\cdot
(d+1) | d<c\}$. Then, $c^+ \cup \{ c\cdot (c+1)\} \cup c^-$ is a
partition of $V$. We show that when proposition (i) is false,
proposition (ii) is necessarily true. Assume that proposition (i)
is false. Since $\mathcal{C}$ contains vertices in both $c^+ \cup
\{c\cdot (c+1)\}$ and $c^-$ (resp. $b$ and $a$), it thus contains
an arc $u\to v$ with $u\in c^-$ and $v\in c^+ \cup \{c\cdot
(c+1)\}$. We must have $u\to v \in F$; if otherwise, $u\to v\in D$
implies $u\to v\in Y$ (since $u< v$), and proposition (i) would be
true, a contradiction. Necessarily, $u = c$ and $v=c\cdot (c+1)$
(because there is no arc in $F$ going out of $c^-$ into $c^+$).
So, $\mathcal{C}$ contains the arc $c \to c\cdot (c+1)$. Using the
same argument, we can show that there is an arc $u^{'} \to v^{'}$
in $\mathcal{C}$ with $u^{'} \in \{c\cdot (c+1)\} \cup c^-$ and
$v^{'} \in c^+$. Since $u^{'} \to v^{'}$ cannot be in $Y$ (since
proposition (i) is false) nor in $X$ (since these arcs go from
$c^+$ to $c^-$), then it must be in $F$, and we can only have
$u^{'} =c\cdot (c+1)$ and $v^{'} = c+1 $. So, $\mathcal{C}$ also
uses the arc $c\cdot (c+1) \to_F c+1$, and thus proposition (ii)
is true. $\qed$ \\

The following two lemmas already appeared verbatim in \cite{BF10},
except that a type II conflict-cycle is additionally imposed here.
However, due to a different definition of conflict-cycles, the
proofs as given in \cite{BF10} are not sufficient \footnote{One
might argue that the corresponding proofs given in \cite{BF10}
shall be sufficient since a type II conflict-cycle is always a
conflict-cycle according to the definition in \cite{BF10}. Note
that, however, a minimal type II conflict-cycle may not be a
minimal conflict-cycle as defined in \cite{BF10}. Therefore, those
proofs are still not sufficient. }.

\begin{lem} \label{lem:404}
Let $\mathcal{C}$ be a minimal type II conflict-cycle where three
vertices $u,e,f\in \Sigma(\mathcal{C})$ are such that
\begin{enumerate}
 \item[-] $\mathcal{C} = u\xrightarrow{P_1}{}^+ e \to_D f
\xrightarrow{P_2}{}^+ u$;
 \item[-] Each of the paths $P_1$ and $P_2$ uses at least one
 vertex from $W$ and at least one arc from $D$.
\end{enumerate}
Then $\eta (e\to_D f) \notin I(u)$.
\end{lem}

\noindent {\em Proof.} \indent (We adapt the proof of Lemma 14 in
\cite{BF10} to our definition of conflict-cycles.) Since
$\mathcal{C}$ is a minimal type II conflict-cycle, by
Lemma~\ref{lem:306}, it must be simple. By contradiction, suppose
that $\eta (e\to_D f) \in I(u)$. Then, by Lemma~\ref{lem:400},
there exists a path $R$ in $D$ connecting either $e$ to $u$ or $u$
to $f$. In the first case, we write $P=P_1$ and $Q=e\to_D f
\xrightarrow{P_2}{}^+ u$, and in the second, $P=P_2$ and
$Q=u\xrightarrow{P_1}{}^+ e \to_D f$, so that there exists a cycle
$\mathcal{C}^{'} = u\xrightarrow{P}{}^+ e \xrightarrow{R}{}_D^* u$
(resp., $\mathcal{C}^{'} = f\xrightarrow{P}{}^+ u
\xrightarrow{R}{}_D^* f$). Since $\mathcal{C}$ is a minimal type
II conflict-cycle, then $R$ cannot be a shortcut, and with $W(Q)$
not being empty, cycle $\mathcal{C}^{'}$ cannot be a
conflict-cycle of type II. Let $W_1(\mathcal{C}^{'}) = \{ i\cdot
(i+1)| a\le i<b \}$. Thus, by Lemma~\ref{lem:114}, for all $c\in
\Sigma(\mathcal{C}^{'})$, we have $c\in [a,b]$, so that
$\Sigma(\mathcal{C}^{'})= \{ i| a\le i\le b\}$ and
$|\mathbb{W}(\mathcal{C}^{'})|=1$. It turns out that
$V(\mathcal{C}^{'}) \subset V(\mathcal{C})$. Note that $R$ does
not use any arc from $F$, so the vertices in
$W_1(\mathcal{C}^{'})$ all come from the path $P$. Moreover,
because the path $P$ is part of the simple conflict-cycle
$\mathcal{C}$ and $|\mathbb{W}(\mathcal{C}^{'})|=1$, the path $P$
(and, the cycles $\mathcal{C}^{'}$ and $\mathcal{C}$ too) must use
a path either $a\to_F b$ or $b\to_F a$. W.l.o.g, this path is
assumed to be $a\to_F b$.

Also note that $P$ uses at least one arc from $D(\mathcal{C})$.
Let $a^{'}\to_D b^{'}$ be such an arc, such that $a^{'}\in
\Sigma(\mathcal{C}^{'})$ and $b^{'}\in \Sigma(\mathcal{C}^{'})$
(i.e., $a\le a^{'} \le b$ and $a\le b^{'}\le b$). If $a^{'} <
b^{'}$, we may write a cycle $\mathcal{C}^{''}=a^{'}\to_D b^{'}
\to_F^* b \to_{E(\mathcal{C})} e \to_D f \xrightarrow{P_2}{}^+ u
\to_{E(\mathcal{C})} a^{'}$, which does not use any vertices in
$W(P_3)$ where the path $P_3=b^{'} \to_{F(P)} a^{'}$. Otherwise,
$a^{'}> b^{'}$, so we may write a cycle $\mathcal{C}^{''}=
a^{'}\to_D b^{'} \to_{F(P)} a^{'}$, which does not use any
vertices in $W(P_2)$. In either case, we can see that
$\mathcal{C}^{''}$ is a subcycle of $\mathcal{C}$, implying that
the latter is not a simple cycle; a contradiction. $\qed$

\begin{lem} \label{lem:406}
Let $\mathcal{C}$ be a minimal type II conflict-cycle, with
$\lambda \ge 5$ joints. Let $e$ and $f$ be two non consecutive
joints of $\mathcal{C}$. Then $\eta(e) \ne \eta(f)$.
\end{lem}

\noindent {\em Proof.} \indent (Please refer to the proof of Lemma
15 in \cite{BF10}, together with Lemma~\ref{lem:404} above.)
$\qed$

  %
  %

\begin{lem} \label{lem:409}
Let $\mathcal{C}$ be a minimal type II conflict-cycle with
$W_1(\mathcal{C}) = \{ i\cdot (i+1) \ | \ a\leq i < b\}$ being an
element of $\mathbb{W}(\mathcal{C})$. Then, we have $a^D\notin
[a,b]$ and $b^D\notin [a,b]$.
\end{lem}

\noindent {\em Proof.} \indent First note that $a\ne a^D$. By
definition, the cycle $\mathcal{C}$ uses an arc from $D$ either
$a\to a^D$ or $a^D\to a$. W.l.o.g., we assume that this arc is
$a\to a^D\in D(\mathcal{C})$. Since $\mathcal{C}$ is a minimal
type II conflict-cycle, it must be simple (by
Lemma~\ref{lem:306}). Moreover, $W_1(\mathcal{C}) = \{ i\cdot
(i+1) \ | \ a\leq i < b\} \in \mathbb{W}(\mathcal{C})$ implies
that $\mathcal{C}$ uses a path either $a\to_F^+b$ or $b\to_F^+ a$.
In the current case, however, this path can only be $b\to_F^+ a$
since $\mathcal{C}$ uses the arc $a\to a^D$ too.

By contradiction, assume that $a^D \in [a,b]$; further, $a <
a^D\le b$ since $a\ne a^D$. It hence implies that there exists a
path $a^D\to_F^+ a$ in $\mathcal{C}$. We may write a cycle
$\mathcal{C}^{'}= a\to a^D\to_F^+ a$, for which any vertex $e\in
\Sigma(\mathcal{C}^{'})$ is such that $a\le e\le b$. On the other
hand, by Lemma~\ref{lem:114}, there exists a vertex $c\in
\Sigma(\mathcal{C})$ such that $c\notin [a,b]$. Thus, $c\notin
\Sigma(\mathcal{C}^{'})$, so that $\mathcal{C}^{'}$ is a subcycle
of $\mathcal{C}$. It however contradicts the fact that
$\mathcal{C}$ is a simple cycle. This proves $a^D\notin [a,b]$. By
using the same arguments above, we can also prove $b^D\notin
[a,b]$. $\qed$

\begin{lem} \label{lem:410}
Let $\mathcal{C}$ be a minimal type II conflict-cycle with
$|\mathbb{W}(\mathcal{C})|\ge 2$ and $W_1(\mathcal{C}) = \{ i\cdot
(i+1) \ | \ a\leq i < b\}$ being an element of
$\mathbb{W}(\mathcal{C})$. Let $c$ be a vertex in $\Sigma$.
\begin{enumerate}
 \item[(i)] If $a<c\le b$ and $\eta(a) \in I(c)$, then $a^D$ and $c$ appear in the same block of the genetic map $\eta (a)$.
 \item[(ii)] If $a\le c<b$ and $\eta(b) \in I(c)$, then $b^D$ and $c$ appear in the same block of the genetic map $\eta (b)$.
\end{enumerate}
\end{lem}

\noindent {\em Proof.} \indent We present below the proof of (i)
only, because (ii) can be proved similarly. Since
$W_1(\mathcal{C}) = \{ i\cdot (i+1) \ | \ a\leq i < b\}$, the
cycle $\mathcal{C}$ uses either the path $a\to_F b$ or $b\to_F a$.
W.l.o.g., we assume that $\mathcal{C}$ uses the path $a\to_F b$.
Because $a<c\le b$, this path goes via the vertex $c$. Since
$\mathcal{C}$ is a minimal type II conflict-cycle, by
Lemma~\ref{lem:409}, we have $a^D \notin [a,b]$. Moreover, by
definition, $\mathbb{W}(\mathcal{C})$ shall contain another
element $W_2(\mathcal{C}) = \{ i\cdot (i+1) \ | \ a^{'}\leq i <
b^{'}\}$, where both vertices $a^{'}$ and $b^{'}$ shall be located
on the path $b^D\to a^D$. W.l.o.g., we assume that $a^{'}$ is
visited before $b^{'}$ in the path $b^D\to a^D$. Thus, we may
write $P$ the path $a^D\to_D a\to_F c$ and $Q$ the path $c\to_F^*
b \to_D b^D \to^* a^{'} \to_F^+ b^{'} \to^* a^D$.

Since $\eta(a) \in I(c)$, $a^D$ and $c$ (and $a$ as well) appear
in the same genetic map numbered $\eta (a)$. So, we distinguish
three cases below.
\begin{enumerate}
 \item[-] In the first case, there exists the path $R=a^D\to_D c$ in $(\Sigma,
 D)$. Let $\mathcal{C}^{'} = c\xrightarrow{Q}{} a^D
 \xrightarrow{R}{} c$. Note that no vertex in $W$ appears in $R$,
 so $W_2(\mathcal{C}) = \{ i\cdot (i+1) \ | \ a^{'}\leq i <
 b^{'}\}$ must appear as an element of $\mathbb{W}(\mathcal{C}^{'})$.
 By Lemma~\ref{lem:106}, we have $b \notin
 [a^{'},b^{'}]$. Then, by Lemma~\ref{lem:114},
 $\mathcal{C}^{'}$ is a conflict-cycle of type II. With $W(P)$ not being
 empty, it follows that $R$ is a shortcut of $\mathcal{C}$, a
 contradiction.
 \item[-] In the second case, there exists the path $R=c\to_D a^D$ in $(\Sigma,
 D)$. Let $\mathcal{C}^{'} = c\xrightarrow{R}{} a^D
 \xrightarrow{P}{} c$. Note that no vertex in $W$ appears in $R$,
 so $W_1(\mathcal{C}^{'}) = \{ i\cdot (i+1) \ | \ a\leq i < c\}$ must
 appear as an element of $\mathbb{W}(\mathcal{C}^{'})$.
 By Lemma~\ref{lem:409}, we have $a^D \notin
 [a,b]$, which implies that $a^D \notin [a,c]$. By Lemma~\ref{lem:114},
 $\mathcal{C}^{'}$ is a conflict-cycle of type II. With $W(Q)$ not being
 empty, it follows that $R$ is a shortcut of $\mathcal{C}$, a
 contradiction.
 \item[-] In the third case, $a^D$ and $c$ are incomparable in $(\Sigma, D)$.
 Since they appear in the same genetic map numbered $\eta (a)$, they
 should appear in the same block of this map.
\end{enumerate}
 $\qed$ \\

It can be seen that the proof of the preceding lemma also implies
the following lemma.

\begin{lem} \label{lem:410a}
Let $\mathcal{C}$ be a minimal type II conflict-cycle with
$W_1(\mathcal{C}) = \{ i\cdot (i+1) \ | \ a\leq i < b\}$ being an
element of $\mathbb{W}(\mathcal{C})$. Let $c$ be a vertex in
$\Sigma$.
\begin{enumerate}
 \item[(i)] If $a<c< b$ and $\eta(a) \in I(c)$, then $a^D$ and $c$ appear in the same block of the genetic map $\eta (a)$.
 \item[(ii)] If $a<c<b$ and $\eta(b) \in I(c)$, then $b^D$ and $c$ appear in the same block of the genetic map $\eta (b)$.
\end{enumerate}
\end{lem}

\begin{lem} \label{lem:411}
Let $w=v\cdot (v+1)\in W$. Then, there exists at most one minimal
type I conflict-cycle being considered during the execution of
\textsc{Approx-MBVS} going via $w$.
\end{lem}

\noindent {\em Proof.} \indent By contradiction, assume that
$\mathcal{C}_1$ and $\mathcal{C}_2$  are two minimal type I
conflict-cycles being considered during the execution of
\textsc{Approx-MBVS}, in this order, such that $w\in
W(\mathcal{C}_1) \cap W(\mathcal{C}_2)$. By definition, let
$W(\mathcal{C}_1) = \{ i\cdot (i+1) \ | \ a_1\leq i < b_1\}$ and
$W(\mathcal{C}_2) = \{ i\cdot (i+1) \ | \ a_2\leq i < b_2\}$.
Since $w = v\cdot (v+1)\in W(\mathcal{C}_1) \cap
W(\mathcal{C}_2)$, we have $a_1\leq v < b_1$ and $a_2\leq v <
b_2$. On the other hand, because the vertices $a_1^F = a_1 \cdot
(a_1+1)$ and $b_1^F = (b_1-1) \cdot b_1$ are removed when
$\mathcal{C}_1$ is considered, they cannot appear in
$\mathcal{C}_2$ so that $a_1<a_2$ and $b_1
> b_2$. Thus, $a_1<a_2<b_2< b_1$, so
that $W(\mathcal{C}_2)$ has a strict subset of $W(\mathcal{C}_1)$.
This, however, contradicts the fact that $\mathcal{C}_1$ is a
minimal conflict-cycle. $\qed$

\begin{lem} \label{lem:411a}
Let $w=v\cdot (v+1)\in W$ and $m=1$. Then, there exists at most
one minimal (type I or type II) conflict-cycle being considered
during the execution of \textsc{Approx-MBVS} going via $w$.
\end{lem}

\noindent {\em Proof.} \indent By contradiction, assume
$\mathcal{C}_1$ and $\mathcal{C}_2$  are two minimal
conflict-cycles being considered during the execution of
\textsc{Approx-MBVS}, in this order, such that $w\in
W(\mathcal{C}_1) \cap W(\mathcal{C}_2)$. By definition, let
$W(\mathcal{C}_1) = \{ i\cdot (i+1) \ | \ a_1\leq i < b_1\}$ and
$W(\mathcal{C}_2) = \{ i\cdot (i+1) \ | \ a_2\leq i < b_2\}$.
Since $w = v\cdot (v+1)\in W(\mathcal{C}_1) \cap
W(\mathcal{C}_2)$, we have $a_1\leq v < b_1$ and $a_2\leq v <
b_2$. On the other hand, because the vertex $a_1^F = a_1 \cdot
(a_1+1)$ is removed when $\mathcal{C}_1$ is considered, $a_1$
cannot appear in $\mathcal{C}_2$ so that $a_1<a_2$. Thus,
$a_1<a_2\le v < b_1$.

By Lemma~\ref{lem:411}, $\mathcal{C}_1$ can only be of type II. By
Lemma~\ref{lem:410}, we further know that
$|\mathbb{W}(\mathcal{C}_1)| = 1$ (since $a_1< v< b_1$). So, the
vertex $b_1^F=(b_1-1)\cdot b_1$ will be removed too when
$\mathcal{C}_1$ is considered. Hence, $b_2 < b_1$, so that $a_1<
a_2 \le v < b_2 < b_1$.

Next we show that there exists a path $u\to_D v$ such that $u\in
[a_2,b_2]$ and $v\in [a_2,b_2]$. To this end, we distinguish two
cases. In the first case, $\mathcal{C}_2$ is assumed to be of type
I. By definition of the type I conflict-cycles, there must exist a
desired path since $\Sigma(\mathcal{C}_2) =\{ i| a_2\le i\le
b_2\}$. In the second case, $\mathcal{C}_2$ is assumed to be of
type II. If $\mathcal{C}_2$ uses the arc $a_2\to a_2^D$, then
there must exist a path $a_2\to_D b_2$. Otherwise, $\mathcal{C}_2$
uses the arc $a_2^D\to a_2$, then there must exist a path
$b_2\to_D a_2$. So, we can always find a path $u\to_D v$ such that
$u\in [a_2,b_2]$ and $v\in [a_2,b_2]$, regardless of the type of
$\mathcal{C}_2$. We further obtain $a_1 < u< b_1$ and $a_1 < v<
b_1$, since $a_1<a_2\le v < b_2 < b_1$. By applying
Lemma~\ref{lem:410a} with $(\mathcal{C}_1, u)$ and
$(\mathcal{C}_1, v)$ successively, we obtain
\begin{enumerate}
 \item[-] $a_1^D$ and $u$ appear in the same block of the only genetic map,
 \item[-] $a_1^D$ and $v$ appear in the same block of the only genetic map.
\end{enumerate}
Therefore, $u$ and $v$ both come from the same block. However, the
existence of the path $u\to_D v$ instead implies that they shall
not appear in the same block, a contradiction. $\qed$

\begin{lem} \label{lem:412}
Let $w=v\cdot (v+1)\in W$, $\mathcal{C}_1$, $\mathcal{C}_2$ and
$\mathcal{C}_3$ three minimal (either type I or type II)
conflict-cycles being considered during the execution of
\textsc{Approx-MBVS}, in this order, such that $w\in \mathcal{C}_1
\cap \mathcal{C}_2 \cap \mathcal{C}_3$. Denote respectively by
$a_1$, $a_2$ and $a_3$ the low joints associated to $w$ in
$\mathcal{C}_1$, $\mathcal{C}_2$ and $\mathcal{C}_3$. Then we
cannot have $\eta(a_1)= \eta(a_2) =\eta(a_3)$.
\end{lem}

\noindent {\em Proof.} \indent By lemma~\ref{lem:411},
$\mathcal{C}_1$ and $\mathcal{C}_2$ must be conflict-cycles of
type II, whereas $\mathcal{C}_3$ could be of either type I or type
II.

By contradiction, assume that $\eta = \eta(a_1) = \eta(a_2)
=\eta(a_3)$. Vertices $a_1$, $a_2$ and $a_3$ are low joints
associated to $w=v\cdot (v+1)$, so $a_1\le v$, $a_2\le v$ and
$a_3\le v$. The vertex $a_1^F = a_1 \cdot (a_1+1)$ is removed when
$\mathcal{C}_1$ is considered, so it cannot appear in
$\mathcal{C}_2$ or $\mathcal{C}_3$. Thus, $a_1<a_2$ and $a_1
<a_3$. Similarly, we can have $a_2<a_3$. Let $W_1(\mathcal{C}_1) =
\{ i\cdot (i+1) \ | \ a_1\leq i < b_1\}$ (resp.,
$W_1(\mathcal{C}_2) = \{ i\cdot (i+1) \ | \ a_2\leq i < b_2\}$) be
the element of $\mathbb{W}(\mathcal{C}_1)$ (resp.,
$\mathbb{W}(\mathcal{C}_2)$) that contains $w=v\cdot (v+1)$. Thus,
$v< b_1$ and $v< b_2$, so $a_2< b_1$, $a_3< b_1$ and $a_3< b_2$.
Then, we may apply Lemma~\ref{lem:410a} with $(\mathcal{C}_1,
a_2)$, $(\mathcal{C}_1, a_3)$ and $(\mathcal{C}_2, a_3)$
successively to obtain
\begin{enumerate}
 \item[-] $a_1^D$ and $a_2$ appear in the same block of genetic map
 $\eta$,
 \item[-] $a_1^D$ and $a_3$ appear in the same block of genetic map
 $\eta$,
 \item[-] $a_2^D$ and $a_3$ appear in the same block of genetic map
 $\eta$.
\end{enumerate}
Therefore, $a_2$ and $a_2^D$ both come from the same block of
genetic map $\eta$, which contradicts $\eta(a_2) = \eta$ (in the
genetic map $\eta(a_2)$, $a_2$ and $a_2^D$ appear in consecutive
blocks). $\qed$

\begin{lem} \label{lem:412a}
Let $w=v\cdot (v+1)\in W$, $\mathcal{C}_1$ and $\mathcal{C}_2$ two
minimal conflict-cycles being considered during the execution of
\textsc{Approx-MBVS}, in this order, such that $w\in \mathcal{C}_1
\cap \mathcal{C}_2$ and $|W(\mathcal{C}_1)|\ge 2$. Denote
respectively by $a_1$ and $a_2$ the low joints associated to $w$
in $\mathcal{C}_1$ and $\mathcal{C}_2$, and by $b_1$ the other
joint (rather than $a_1$) associated to $w$ in $\mathcal{C}_1$.
Then we cannot have $\eta(a_1)=\eta(b_1) = \eta(a_2)$.
\end{lem}

\noindent {\em Proof.} \indent By lemma~\ref{lem:411},
$\mathcal{C}_1$ must be a conflict-cycle of type II, whereas
$\mathcal{C}_2$ could be of either type I or type II.

By contradiction, assume that $\eta = \eta(a_1) =\eta(b_1) =
\eta(a_2)$. As shown in the preceding lemma, we have $a_1< a_2 \le
v < b_1$. Then, we may apply Lemma~\ref{lem:410} to obtain
\begin{enumerate}
 \item[-] $a_1^D$ and $a_2$ appear in the same block of genetic map
 $\eta$,
 \item[-] $b_1^D$ and $a_2$ appear in the same block of genetic map
 $\eta$,
 \item[-] $a_1$ and $b_1^D$ appear in the same block of genetic map
 $\eta$.
\end{enumerate}
Therefore, $a_1$ and $a_1^D$ both come from the same block of
genetic map $\eta$, which contradicts $\eta(a_1) = \eta$ (in the
genetic map $\eta(a_1)$, $a_1$ and $a_1^D$ appear in consecutive
blocks). $\qed$

\begin{lem} \label{lem:414}
Let $w\in W$ and $\mathbb{C}$ the set of all the minimal
conflict-cycles being considered during the execution of
\textsc{Approx-MBVS} going via $w$. Let $J_w$ denote the total
number of joints being selected in these cycles (in order to
remove adjacencies). Then, $J_w\le m^2+2m-1$.
\end{lem}

\noindent {\em Proof.} \indent We write $w=v\cdot (v+1) \in W$,
and $\mathbb{C} = \{ \mathcal{C}_1,\dots, \mathcal{C}_q\}$ the set
of the $q$ conflict-cycles being considered, in this order, during
the execution of \textsc{Approx-MBVS}. In each cycle
$\mathcal{C}_h$, $w$ can be associated to a low joint $v_h$ and to
the corresponding deleted vertex $w_h = v_h^F = v_h \cdot (v_h
+1)$. We write $\lambda_h$ the number of joints of
$\mathcal{C}_h$. If $\mathcal{C}_h$ is a minimal type II
conflict-cycle, then $\frac{\lambda_h}{2}$ is the number of low
joints (and thus the maximum number of deleted vertices) in this
cycle. Otherwise, it is of type I, so $\frac{\lambda_h}{2} = 1$,
but the number of deleted vertices in this cycle could be up to 2.
Since $w_h$ is deleted while $\mathcal{C}_h$ is considered, we
have $w_h\notin W(\mathcal{C}_{h^{'}})$ and $v_h < v_{h^{'}} \le
v$, for all $h^{'}>h$. Indeed, $\forall u\in \{ v_{h^{'}}, \dots,
v \}$, the vertex $u\cdot (u+1)$ belongs to
$W(\mathcal{C}_{h^{'}})$.

By Lemma~\ref{lem:411}, there exists at most one minimal type I
conflict-cycle being considered during the execution of
\textsc{Approx-MBVS} going via $w$. Thus, the first $q-1$ cycles
must be all of type II, while the last cycle $\mathcal{C}_q$ may
be of either type I or type II, depending on whether a minimal
type I conflict-cycle is considered or not.

Consider now the list $\langle\eta(v_1),\eta(v_2), \cdots,
\eta(v_q)\rangle$. Unlike in a set, duplicate values are allowed
in a list. By Lemma~\ref{lem:412}, we know that no value can
appear more than twice in the list. Hence, $q\le 2m$. Indeed, we
can further show below that $q\le 2m-1$ when $\lambda_1 \ge 4$
(i.e., when $|\mathbb{W}(\mathcal{C}_1)| \ge 2$). By
contradiction, suppose that $q=2m$ when $\lambda_1 \ge 4$. So,
$q\ge 2$, which implies that there are at least two minimal
conflict-cycles being considered during the execution of
\textsc{Approx-MBVS} going via $w$. By Lemma~\ref{lem:411}, the
first conflict-cycle $\mathcal{C}_1$ must be of type II. Let $e$
be the other joint rather than $v_1$ in $\mathcal{C}_1$ associated
to $w$. Because $q= 2m$, by Lemma~\ref{lem:412}, we can find
exactly two distinct vertices $v_i$ and $v_j$ such that $\eta(e) =
\eta(v_i) = \eta(v_j)$ and $1\le i <j\le q=2m$. Recall that $v_i$
and $v_j$ are the respective low joints of $\mathcal{C}_i$ and
$\mathcal{C}_j$ that are both associated to $w$. So, neither $v_i$
nor $v_j$ coincide with $e$ (but $v_i$ might coincide with $v_1$)
and, moreover, $v_1 \le v_i < v_j < e$. By using
Lemma~\ref{lem:410} with $(\mathcal{C}_1, v_i)$, $(\mathcal{C}_1,
v_j)$ and $(\mathcal{C}_i, v_j)$ successively, we obtain
\begin{enumerate}
 \item[-] $e^D$ and $v_i$ appear in the same block of genetic map
 $\eta$,
 \item[-] $e^D$ and $v_j$ appear in the same block of genetic map
 $\eta$,
 \item[-] $v_i^D$ and $v_j$ appear in the same block of genetic map
 $\eta$.
\end{enumerate}
It turns out that both $v_i$ and $v_i^D$ come from the same block
of genetic map $\eta$, which contradicts the fact that $v_i$ and
$v_i^D$ shall appear in consecutive blocks. So, this proves that
$q\le 2m-1$ when $\lambda_1 \ge 4$.

Consider now the list $\langle\eta(v_{h+1}),\eta(v_{h+2}), \cdots,
\eta(v_q)\rangle$. Let $m_1$ and $m_2$ denote respectively the
number of unique values and the number of duplicate values in the
above list (duplicated values being counted only once). By
Lemma~\ref{lem:412}, we know that no value can appear more than
twice in the list. Then, we obtain the following equation.
\begin{equation}
m_1 + 2 m_2 = q-h. \label{eqn:104}
\end{equation}

Let us assume for a moment that $\lambda_h \ge 5$, i.e.,
$\mathcal{C}_h$ has more than four joints. Let $e_1$, $e_2$, $e_3$
and $e_4$ be four joints such that $\mathcal{C}_h$ uses the path
$e_1\to_D^+ e_2\to_F^+ w \to_F^+ e_3\to_D^+ e_4$. Note that either
$e_2 = v_h$ or $e_3 = v_h$. And, for all $h^{'} > h$, the vertex
$v_{h^{'}}$ appears between joints $e_2$ and $e_3$, so we may
write $\mathcal{C} = e_1\to_D^+ e_2\to_F^+ v_{h^{'}} \to_F^+
e_3\to_D^+ e_4 \to^+ e_1$. Consider a joint $e_i$ rather than
$e_1$, $e_2$, $e_3$ and $e_4$, for all $i\in [5, \lambda_h]$. We
have either $\mathcal{C} = e_1\to_D^+ e_2\to_F^+ v_{h^{'}} \to_F^+
e_3\to_D^+ e_4 \to^+ e_i \to_D e_i^D \to e_1$ or $\mathcal{C} =
e_1\to_D^+ e_2\to_F^+ v_{h^{'}} \to_F^+ e_3\to_D^+ e_4 \to^+ e_i^D
\to_D e_i \to e_1$. In either case, using Lemma~\ref{lem:404} with
three vertices $v_{h^{'}}$, $e_i$ and $e_i^D$, we have $\eta(e_i)
\notin I(v_{h^{'}})$, for all $i\in [5, \lambda_h]$ and all
$h^{'}> h$. In other words, for each value $\eta$ counted into
$m_1$ or $m_2$, we cannot have any joint $e_i$ for $i\in [5,
\lambda_h]$ such that $\eta = \eta(e_i)$.

Consider now the list $\langle\eta(e_1),\eta(e_2), \eta(e_3),
\eta(e_4)\rangle$. Let $m_3$ and $m_4$ denote the number of values
(duplicated values being counted only once) in this list that
appear or do not appear in the preceding list
$\langle\eta(v_{h+1}), \eta(v_{h+2}), \dots, \eta(v_q)\rangle$,
respectively. First, note that $e_1$ and $e_3$ are two non
consecutive joints of $\mathcal{C}_h$. By Lemma~\ref{lem:406}, we
cannot have $\eta(e_1)= \eta(e_3)$, which implies that
\begin{equation}
m_3+m_4\ge 2. \label{eqn:108}
\end{equation}
Then, consider each value $\eta$ counted into $m_2$. By definition
of $m_2$, we have two distinct vertices $v_i$ and $v_j$ such that
$\eta = \eta(v_i) = \eta(v_j)$ and $h< i<j \le q$. By using the
same arguments above as in the preceding paragraph, we can show
that this value $\eta$ won't be counted into $m_3$. It hence
follows that
\begin{equation}
m_3\le m_1. \label{eqn:110}
\end{equation}
In addition, for each value $\eta$ counted into $m_4$, by
Lemma~\ref{lem:406}, we cannot have two distinct joints $e_i$ and
$e_j$ for $i,j\in [5, \lambda_h]$ such that $\eta = \eta(e_i) =
\eta(e_j)$.

To summarize, for each value $\eta$ counted into $m_1$ or $m_2$,
there is no joint $e_i$ for $i\in [5, \lambda_h]$ such that $\eta
= \eta(e_i)$. For each value $\eta$ counted into $m_4$, there
exists at most one joint $e_i$ for $i\in [5, \lambda_h]$ such that
$\eta = \eta(e_i)$. For any other possible value $\eta$, there
exist at most two joints $e_i$ and $e_j$ for $i,j\in [5,
\lambda_h]$ such that $\eta = \eta(e_i) = \eta(e_j)$; moreover,
the total of such possible $\eta$ values (i.e., all the $\eta$
values attained by the joints other than $e_1$, $e_2$, $e_3$ and
$e_4$) is no more than $m-m_1-m_2-m_4$. Based on these
observations, we can deduce the following inequality:
\begin{equation}
\lambda_h - 4 \le 2(m - m_1 - m_2 - m_4) + m_4. \label{eqn:112}
\end{equation}
Note that $\lambda_h$ is always even. Then, by using the above
Equality~\ref{eqn:104} and Inequalities~\ref{eqn:108},
~\ref{eqn:110}, and~\ref{eqn:112}, we obtain the following
inequality for $\lambda_h \ge 5$:
\begin{equation}
\frac{\lambda_h}{2} \le  m- \left\lceil \frac{q-h}{2} \right\rceil
+ 1. \label{eqn:114}
\end{equation}
This inequality also holds when $\lambda_h =2$ because $q\le 2m$
and $h\ge 1$. When $\lambda_h =4$, it does not hold only when $q =
2m$ and $h=1$. However, this condition will never be met because
we have shown above that $q\le m-1$ when $\lambda_1 = 4$. To
summarize, the above inequality holds for all $\lambda_h \ge 2$,
although it is initially derived based on the assumption that
$\lambda_h \ge 5$. Further note that the above inequality holds
for all $m\ge 1$.

Let us assume for a moment that $m \ge 2$. By Lemma~\ref{lem:406},
we have that $\lambda_h \le 2m$ when $m\ge 2$. Thus,
$\frac{\lambda_h}{2} \le \min \left(m, m-\left\lceil \frac{q-h}{2}
\right\rceil + 1 \right)$ holds for all the conflict-cycles being
considered during the execution of \textsc{Approx-MBVS},
regardless of their types.

Recall that, for a possible minimal type I conflict-cycle
$\mathcal{C}_q$, the algorithm will select two joints rather than
one joint (as computed by $\frac{\lambda_q}{2}$). By incorporating
this, we then obtain (assume that $m\ge 2$)
\begin{equation}
\begin{array}{ccl}
  J_w & =   & \max\{2,  \frac{\lambda_q}{2} \} + \sum\limits_{h=1}^{q-1} \frac{\lambda_h}{2} \\ \\
      & \le & m +\sum\limits_{h=1}^{q-1} \min \left(m, m-\left\lceil \frac{q-h}{2} \right\rceil + 1 \right) \\ \\
      & = & m + \sum\limits_{h=1}^{q-1} \left(m-\left\lceil \frac{h}{2} \right\rceil + 1 \right) \\ \\
      & \le & m + \sum\limits_{h=1}^{2m-1} \left(m-\left\lceil \frac{h}{2} \right\rceil + 1 \right) \\ \\
      & = & m + \sum\limits_{h=1}^{2m-1} (m + 1) - \sum\limits_{h=1}^{2m-1} \left\lceil \frac{h}{2} \right\rceil  \\ \\
      & = & 2m^2 + 2m -1 - \left(m+2\sum\limits_{h=1}^{m-1} h\right)  \\ \\
      & = & m^2 + 2m -1.
\end{array} \nonumber
\end{equation}

In case of $m=1$, by Lemma~\ref{lem:411a}, we have $q=1$ (we
assume here that at least one conflict-cycle being considered
going via $w$; otherwise, $J_w =0$). No matter whether this cycle
$\mathcal{C}_1$ is of type I, of type II with
$|\mathbb{W}(\mathcal{C}_1)|=1$, or of type II with
$|\mathbb{W}(\mathcal{C}_1)|\ge 2$, the algorithm will select
exactly two joints only, thereby making $J_w\le m^2 + 2m -1$ still
true. In conclusion, $J_w\le m^2 + 2m -1$ holds for all $m\ge 1$.
$\qed$

\begin{cor} \label{cor:416}
Let $w\in W$ and $\mathbb{C}$ the set of all the conflict-cycles
being considered during the execution of \textsc{Approx-MBVS}
going via $w$. Then, the total number of vertices in $W$ to be
removed from cycles of $\mathbb{C}$ is bounded from the above by
$m^2 + 2m -1$.
\end{cor}

\begin{thm} \label{thm:MBVS}
Algorithm \textsc{Approx-MBVS} achieves an $(m^2 + 2m
-1)$-approximation for the \textsc{MBVS} problem, where $m$ is the
number of genetic maps used to create the input adjacency-order
graph.
\end{thm}

\noindent {\em Proof.} \indent Correctness of Algorithm
\textsc{Approx-MBVS} follows from Corollary~\ref{cor:416}, since
the algorithm removes at least one vertex from each
conflict-cycle. Let $W^o=\{ w_1^o,\dots, w_k^o\}$ be an optimal
solution of size $k$, i.e., a minimum breakpoint vertex set of
$G_\Pi$. For each $w_i^o$, the algorithm deletes at most $(m^2 +
2m -1)$ adjacencies of $W$ (by Corollary~\ref{cor:416}). Since
every cycle being considered by the algorithm goes through some
$w_i^o$, the total size of the output solution is at most $k\cdot
(m^2 + 2m -1)$. The next subsection shows that the algorithm can
be executed in polynomial time. $\qed$

\subsection{Running time} \label{sec:time}

The remaining question in the algorithm \textsc{Approx-MBVS} is
whether there exists any polynomial-time algorithm to find a
minimal conflict-cycle from an induced subgraph $G_\Pi[W^{'}\cup
\Sigma]$. Since the algorithm considers all the type II
conflict-cycles before any type I conflict-cycle is considered, we
present first the algorithm to find a minimal conflict-cycle of
type II in the below.

\subsubsection{Finding a minimal type II conflict-cycle}

First of all, we can develop a procedure to determine whether a
given cycle is a conflict-cycle (following the definition) and, if
it is, further determine whether it is of type I or of type II
(following Lemma~\ref{lem:114}). We denote this procedure by
CC$_{II}$-check$()$, and note that it can be executed in $O(n)$
time.

\begin{lem} \label{lem:500}
Let $W^{'}$ be a subset of $W$. If $G_\Pi[W^{'}\cup \Sigma]$
contains a type II conflict-cycle, then it also contains a type II
conflict-cycle $\mathcal{C} = a \xrightarrow{P}{} c
\xrightarrow{Q}{} b \to_F^+ a$ such that (i) $a,b,c\in \Sigma$,
(ii) neither $a\le c \le b$ nor $b\le c\le a$, and (iii) $P$ and
$Q$ are the respective shortest paths between two vertices in the
induced subgraph $G_\Pi[W^{''}\cup \Sigma]$ where $W^{''}=W^{'} -
\{ (a-1)\cdot a, a\cdot (a+1), (b-1)\cdot b, b\cdot (b+1) \}$.
\end{lem}

\noindent {\em Proof.} \indent Since $G_\Pi[W^{'}\cup \Sigma]$
contains a conflict-cycle of type II, by Lemma~\ref{lem:118}, it
also contains a simple conflict-cycle of type II. Let this simple
conflict-cycle be $\mathcal{C}^{'}$, with $W_1(\mathcal{C}^{'})=\{
i\cdot (i+1) | a_1\le i < b_1 \}$. By Lemma~\ref{lem:114}, there
exists a vertex $c\in \Sigma(\mathcal{C}^{'})$ such that $c\notin
[a_1,b_1]$. So, we have either $\mathcal{C}^{'} = a_1\to c\to b_1
\to_F^+ a_1$ or $\mathcal{C}^{'} = a_1 \to_F^+ b_1\to c\to a_1$.
In the first case, we let $a=a_1$ and $b=b_1$; in the second case,
let $a=b_1$ and $b=a_1$. In both cases, $\mathcal{C}^{'}$ uses the
path $R = a \to c \to b$.

Recall that $\mathcal{C}^{'}$ is simple, so $R$ won't traverse any
vertices from the set $\{ (a-1)\cdot a, a\cdot (a+1), (b-1)\cdot
b, b\cdot (b+1) \}$. It turns out that the path $R$ is fully
contained in the induced subgraph $ G_\Pi[W^{''}\cup \Sigma]$
where $W^{''}=W^{'} - \{ (a-1)\cdot a, a\cdot (a+1), (b-1)\cdot b,
b\cdot (b+1) \}$. Since there exists in $ G_\Pi[W^{''}\cup
\Sigma]$ an path from $a$ to $c$ and also an path from $c$ to $b$,
we may write their respective shortest paths $a \xrightarrow{P}{}
c$ and $c \xrightarrow{Q}{} b$. Thus, we obtain a new cycle
$\mathcal{C} = a \xrightarrow{P}{} c \xrightarrow{Q}{} b \to_F^+
a$. Note that the path $a \xrightarrow{P}{} c \xrightarrow{Q}{} b$
could not traverse any vertex from the set $\{ (a-1)\cdot a,
a\cdot (a+1), (b-1)\cdot b, b\cdot (b+1) \}$, so that $\{ i\cdot
(i+1) | a_1\le i < b_1 \}$ is also an element of
$\mathbb{W}(\mathcal{C})$ and, moreover, $c \notin [a_1, b_1]$. It
hence follows from Lemma~\ref{lem:114} that $\mathcal{C}$ is a
conflict-cycle of type II. $\qed$ \\

Based on the above lemma, we propose a procedure to determine
whether a given graph $G_\Pi[W^{'}\cup \Sigma]$ contains a type II
conflict-cycle and, if any, to report one. It is done by
conducting four tests for all triples of distinct vertices
$\langle a, b, c\rangle \in \Sigma \times \Sigma \times \Sigma$:
(i) whether $c\notin [a,b]$ if $a<b$ and $c\notin [b,a]$ if $b<a$
(taking $O(n)$ time), (ii) whether there exist all the vertices of
$\{ i\cdot (i+1) | a\le i < b\ {\rm or}\ b\le i <a\}$ in
$G_\Pi[W^{'}\cup \Sigma]$ (taking $O(n)$ time), (iii) whether
there exists a shortest path $a \xrightarrow{P}{} c$ between $a$
and $c$ in $G_\Pi[W^{''}\cup \Sigma]$ (taking $O(n^2)$ time), and
(iv) whether there exists a shortest path $c \xrightarrow{Q}{} b$
between $c$ and $b$ in $G_\Pi[W^{''}\cup \Sigma]$ (taking $O(n^2)$
time). If a triple $\langle a, b, c\rangle$ passes all the four
tests, then we find a type II conflict-cycle $\mathcal{C} = a
\xrightarrow{P}{} c \xrightarrow{Q}{} b \to_F a$. If, instead, no
triples in $\Sigma \times \Sigma \times \Sigma$ can pass them,
then we know that $G_\Pi[W^{'}\cup \Sigma]$ contains no
conflict-cycles of type II. We denote this procedure by
CC$_{II}$-seed$()$, and note that it can be executed in time
$O(n^5)$.

It is worth noting that the conflict-cycle $\mathcal{C}$ found by
the above procedure CC$_{II}$-seeding$()$ is not necessarily
simple. If $\mathcal{C}$ is not simple, by Lemma~\ref{lem:118} we
know that there must exist a simple type II conflict-subcycle of
$\mathcal{C}$. To find it, we propose a procedure, called
CC$_{II}$-simplify$()$, which works by mainly applying
CC$_{II}$-check$()$ to every simple subcycle of $\mathcal{C}$.
Note that the procedure CC$_{II}$-simplify$()$ can also be
executed in $O(n)$ time.

By applying the procedures CC$_{II}$-seed$()$ and
CC$_{II}$-simplify$()$ successively, we may obtain a simple type
II conflict-cycle (if any) from $G_\Pi[W^{'}\cup \Sigma]$. The
next lemma then tells us how to find a minimal conflict-cycle of
type II.

\begin{lem} \label{lem:502}
Let $\mathcal{C}$ be a simple conflict-cycle of type II. If it has
a shortcut, then it also contains a shortcut $R=u
\xrightarrow{R_1}{}_D^* w \xrightarrow{R_2}{}_D^* v$ such that (i)
$u, v\in \Sigma(\mathcal{C})$, (ii) $w\in \Sigma$, and (iii) $R_1$
and $R_2$ are the respective shortest paths between two vertices
in $(\Sigma, D)$.
\end{lem}

\noindent {\em Proof.} \indent Since $\mathcal{C}$ has a shortcut,
let this shortcut be the path $u \xrightarrow{R^{'}}{}_D^+ v$
(note that $u\ne v$ because $\mathcal{C}$ is simple). By
definition, we know that (i) $u,v\in \Sigma(\mathcal{C})$ , so we
may write $\mathcal{C} = v \xrightarrow{P}{}^+ u
\xrightarrow{Q}{}^+ v$, (ii) the cycle $\mathcal{C}^{'} = v
\xrightarrow{P}{}^+ u \xrightarrow{R^{'}}{}_D^* v$ is also a
conflict-cycle of type II, and (iii) $W(Q)\ne \emptyset$.

Let $W_1(\mathcal{C}^{'})=\{ i\cdot (i+1) | a_1\le i < b_1 \}$.
Since $\mathcal{C}^{'}$ is a conflict-cycle of type II, by
Lemma~\ref{lem:114}, there exists a vertex $w^{'} \in
\Sigma(\mathcal{C}^{'})$ such that $w^{'} \notin [a_1, b_1]$. If
$w^{'}$ is located on the path $P$, then let $w=b$; otherwise,
$w^{'}$ is located on the path $R^{'}$, and we instead let
$w=w^{'}$. We can see that, in both cases, there exits in
$(\Sigma, D)$ at least one path from $u$ to $w$ and also at least
one path from $w$ to $v$. Let $u \xrightarrow{R_1}{}_D^* w$ and $w
\xrightarrow{R_2}{}_D^* v$ denote their respective shortest paths,
so we may write the path $R=u \xrightarrow{R_1}{}_D^* w
\xrightarrow{R_2}{}_D^* v$. Thus, we obtain a new cycle
$\mathcal{C}^{''} = v \xrightarrow{P}{}^+ u \xrightarrow{R}{}_D^*
v$. To show $R$ is a shortcut of $\mathcal{C}$, it suffices by
showing that the cycle $\mathcal{C}^{''}$ is a conflict-cycle of
type II, as done below.

Note that $W(\mathcal{C}^{'}) = W(\mathcal{C}^{''})$, since
neither $R$ nor $R^{'}$ use any vertex from $W$. Consequently,
$\mathbb{W}(\mathcal{C}^{'}) = \mathbb{W}(\mathcal{C}^{''})$,
which implies that $\{ i\cdot (i+1) | a_1\le i < b_1 \}$ is also
an element of $ W(\mathcal{C}^{''})$. Further note that, no matter
in which case the vertex $w$ is defined, the vertex $w^{'}$ is
always in $\Sigma(\mathcal{C}^{''})$ so that $w^{'} \notin [a_1,
b_1]$. Thus, it follows from Lemma~\ref{lem:114} that
$\mathcal{C}^{''}$ is a conflict-cycle of type II. $\qed$ \\

Based on the above lemma, we propose a procedure \footnote{The
main challenge in developing such a procedure is to ensure that it
would not end up with a conflict-cycle of type I.} to determine
whether a given simple type II conflict-cycle $\mathcal{C}$ is
minimal and, if it is not minimal, to report a type II
conflict-cycle $\mathcal{C}^{'}$ with $W(\mathcal{C}^{'}) <
W(\mathcal{C})$. It is done by conducting four tests for all
triples of vertices $\langle u, v, w\rangle \in
\Sigma(\mathcal{C}) \times \Sigma(\mathcal{C}) \times \Sigma$: (i)
whether $W(Q)\ne \emptyset$ where $\mathcal{C}= v
\xrightarrow{P}{}^+ u \xrightarrow{Q}{}^+ v$ (taking $O(n)$ time),
(ii) whether there exists a shortest path $u
\xrightarrow{R_1}{}_D^* w$ between $u$ and $w$ in $(\Sigma, D)$
(taking $O(n^2)$ time), (iii) whether there exists a shortest path
$w \xrightarrow{R_2}{}_D^* v$ between $w$ and $v$ in $(\Sigma, D)$
(taking $O(n^2)$ time), and (iv) whether the cycle
$\mathcal{C}^{'} = v \xrightarrow{P}{}^+ u \xrightarrow{R_1}{}_D^*
w \xrightarrow{R_2}{}_D^* v$ is a conflict-cycle of type II by
using the procedure CC$_{II}$-check$()$ (taking $O(n)$ time). If a
triple $\langle u,v,w\rangle$ passes all the four tests, then we
find a type II conflict-cycle $\mathcal{C}^{'}$ such that
$W(\mathcal{C}^{'}) < W(\mathcal{C})$ (i.e., the path
$u\xrightarrow{R_1}{}_D^* w \xrightarrow{R_2}{}_D^* v$ is a
shortcut of $\mathcal{C}$). If, instead, no triples in
$\Sigma(\mathcal{C}) \times \Sigma(\mathcal{C}) \times \Sigma$ can
pass them, then we know that $\mathcal{C}$ is already minimal. We
denote this procedure by CC$_{II}$-reduce$()$, and note that it
can be executed in time $O(n^5)$.

We present in Table~\ref{tab:typeII} the algorithm used to find a
minimal type II conflict-cycle from an adjacency-order (sub)graph.
Note that $W(\mathcal{C}^{'}) < W(\mathcal{C})$ holds after each
execution of the {\bf while} loop, so that the {\bf while} loop
cannot be repeated more than $n$ times. Thus, we can see that this
algorithm can be executed in $O(n^6)$ time.

\begin{table}[t] \centering
\begin{tabular}{|l|}
  \hline
   \rowcolor[rgb]{0.9,0.9,0.9} {\bf Algorithm} \textsc{Find-a-Minimal-Type-II-Conflict-Cycle}  \\
  \hline
   {\bf input} An induced adjacency-order subgraph $G_\Pi[W^{'}\cup \Sigma]$ \\
   {\bf output} A minimal type II conflict-cycle $\mathcal{C}$ \\
   {\bf begin} \\
  \ \ \ $\mathcal{C} \leftarrow$ CC$_{II}$-seed$()$; \\
  \ \ \ $\mathcal{C}^{'} \leftarrow \mathcal{C}$; \\
  \ \ \ {\bf while} $\mathcal{C}^{'} \ne \emptyset$  \\
  \ \ \ \quad $\mathcal{C} \leftarrow \mathcal{C}^{'}$; \\
  \ \ \ \quad $\mathcal{C} \leftarrow$ CC$_{II}$-simplify$(\mathcal{C})$; \\
  \ \ \ \quad $\mathcal{C}^{'} \leftarrow$ CC$_{II}$-reduce$(\mathcal{C})$; \\
  \ \ \ {\bf return} $\mathcal{C}$; \\
   {\bf end}  \\
  \hline
\end{tabular} \caption{A polynomial-time algorithm for finding a minimal type II
conflict-cycle from an induced adjacency-order subgraph
$G_\Pi[W^{'}\cup \Sigma]$. Note that $G_\Pi[W^{'}\cup \Sigma] =
G_\Pi$ if $W^{'} = W$. } \label{tab:typeII}
\end{table}

\subsubsection{Finding a minimal type I conflict-cycle}

The algorithm \textsc{Approx-MBVS} starts the search for the
minimal type I conflict-cycle only when there are no longer any
type II conflict-cycles contained in the subgraph $G_\Pi[W^{'}
\cup \Sigma]$. The following lemma assists us in developing an
algorithm to find a minimal type I conflict-cycle from
$G_\Pi[W^{'} \cup \Sigma]$.

\begin{lem} \label{lem:504}
Let $W^{'}$ be a subset of $W$. If $G_\Pi[W^{'}\cup \Sigma]$
contains a type I conflict-cycle, then it also contains a type I
conflict-cycle $\mathcal{C} = a_1 \xrightarrow{e_1}{} b_1 \to_F^*
a_2 \xrightarrow{e_2}{} b_2 \to_F^* a_1$ such that (i) the arcs
$e_1\in X$ and $e_2\in Y$, (ii) $V(\mathcal{C}) = \{ i\cdot (i+1)
\ | \ a\leq i < b\} \cup \{ i \ | \ a\leq i \leq b\}$ where $a
=\min \{a_1, b_1, a_2, b_2 \}$ and $b =\max \{a_1, b_1, a_2, b_2
\}$, and (iii) $D(\mathcal{C}) = \{ e_1, e_2\}$.
\end{lem}

\begin{figure*}[t]
\begin{center}
  \includegraphics[width=1.0\textwidth]{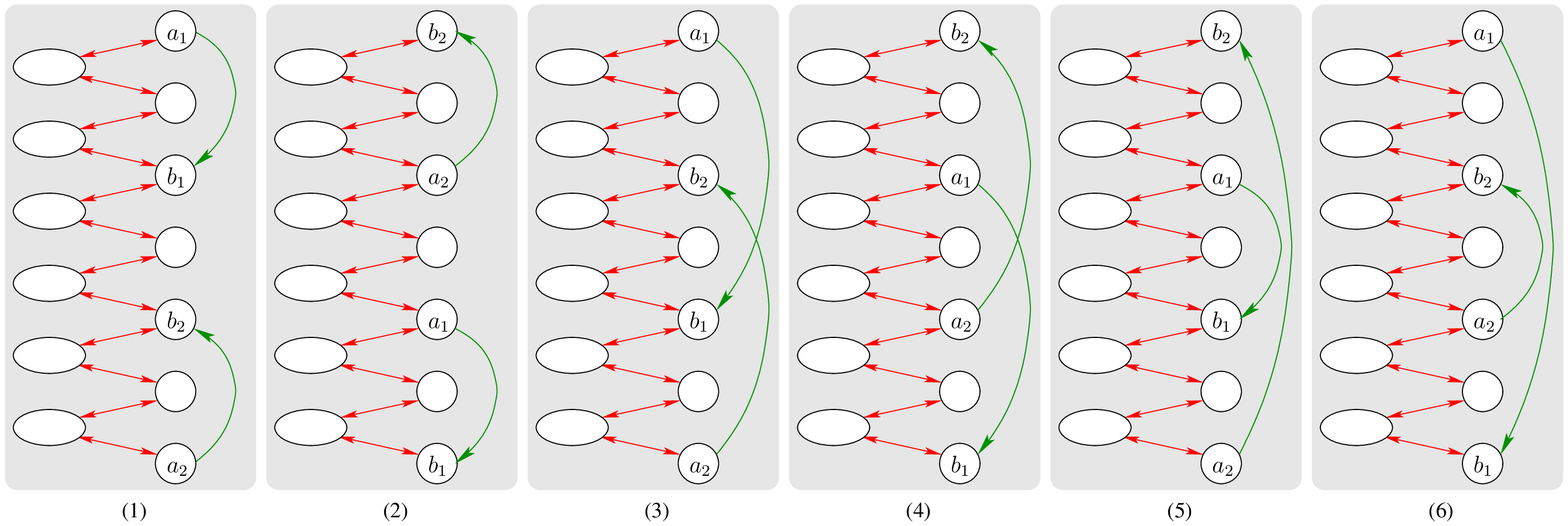}
\caption{A conflict-cycle of type I can be formed for each of the
six general cases as follows: (1)\ \  $a_1\to_X b_1 \to_F^* b_2
\to_F a_2 \to_Y b_2 \to_F^* b_1 \to_F a_1$, \ \ (2)\ \   $a_1\to_X
b_1 \to_F a_1 \to_F^* a_2 \to_Y b_2 \to_F a_2 \to_F^* a_1$, \ \
(3)\ \ $a_1\to_X b_1 \to_F a_2 \to_Y b_2 \to_F^* b_1 \to_F^* b_2
\to_F a_1$, \ \ (4)\ \   $a_1\to_X b_1 \to_F a_2 \to_F^* a_1
\to_F^* a_2 \to_Y b_2 \to_F a_1$, \ \ (5)\ \   $a_1\to_X b_1 \to_F
a_1 \to_F b_1 \to_F^* a_2 \to_Y b_2 \to_F^* a_1$, \ \ (6)\ \
$a_1\to_X b_1 \to_F^* a_2 \to_Y b_2 \to_F a_2 \to_F b_2 \to_F^*
a_1$.} \label{fig:six-cases-of-type-I}
\end{center}
\end{figure*}

\noindent {\em Proof.} \indent Since $G_\Pi[W^{'}\cup \Sigma]$
contains a type I conflict-cycle, by definition, it shall use one
arc $e_1= a_1\to b_1 \in X$, one arc $e_2 = a_2\to b_2\in Y$, and
all the vertices of $\{ i\cdot (i+1) \ | \ a\leq i < b\} \cup \{ i
\ | \ a\leq i \leq b\}$ if we let $a =\min \{a_1, b_1, a_2, b_2
\}$ and $b =\max \{a_1, b_1, a_2, b_2 \}$. With these arcs and
vertices, we are able to construct a desired type I conflict-cycle
$\mathcal{C}$ through a case study, as illustrated in
Figure~\ref{fig:six-cases-of-type-I}. $\qed$ \\

Based on the above lemma, we propose the following algorithm to
find a minimal type I conflict-cycle (if any). For all pairs of
arcs $\langle e_1, e_2 \rangle \in X \times Y$, where $e_1=a_1\to
b_1 \in X$ and $e_2=a_2\to b_2 \in Y$, first compute $a =\min
\{a_1, b_1, a_2, b_2 \}$ and $b =\max \{a_1, b_1, a_2, b_2 \}$ and
then test if there exists a path $a\to_F b$ from $a$ to $b$ using
arcs all from $F$ (each taking $O(n)$ time). Among all those pairs
that passed the test, the one that attains the smallest value of
$(b-a)$ will be returned as a minimal type I conflict-cycle. Note
that this algorithm can be executed in $O(n^5)$ time since the
total number of arc pairs is no more than $O(n^4)$.

Consider now the whole execution of the algorithm
\textsc{Approx-MBVS}. Note that two {\bf while} loops of
\textsc{Approx-MBL} cannot each be repeated more than $n$ times
because we delete at least one vertex in $F$ for each minimal
conflict-cycle $\mathcal{C}$ to be considered. Therefore, the
algorithm \textsc{Approx-MBVS} (and hence Algorithm
\textsc{Approx-MBL}) can be executed in $O(n^7)$ time. The main
result of this paper thus follows (the approximation ratio follows
from Theorem~\ref{thm:MBVS}).

\begin{thm}
Algorithm \textsc{Approx-MBL} achieves an approximation ratio of
$(m^2 + 2m -1)$ for the MBL problem and runs in $O(n^7)$ time,
where $m$ is the number of genetic maps used to create the input
partial order and $n$ the total number of distinct genes appearing
in these maps.
\end{thm}

\section{Conclusions} \label{sec:conclusion}

In this paper, we have studied the MBL problem in its original
version, i.e., it assumes that gene strandedness is not available
in the input genetic maps. We found that the approximation
algorithm proposed in \cite{BF10} for the MBL problem is not
applicable here because it implicitly requires the availability of
gene strandedness. Therefore, we revised the definition of
conflict-cycle in the adjacency-order graphs, and then developed
an approximation algorithm by basically generalizing the algorithm
in \cite{BF10}. It achieves a ratio of $(m^2+2m-1)$ and runs in
$O(n^7)$ time, where $m$ is the number of genetic maps used to
construct the input partial order and $n$ the total number of
distinct genes in these maps. We believe that the same
approximation ratio also applies to the special variant of the MBL
problem studied in \cite{BF10}, thereby achieving an improved
approximation ratio over the previous one $(m^2 +4m-4)$ given in
\cite{BF10}. In the future, it is very interesting to investigate
whether an $O(m)$-approximation can be achieved for the MBL
problem.

%

\bibliography{UMBL}{}
\bibliographystyle{plain}

\end{document}